\documentclass[article,authoryear,aas]{beg_32}             

\usepackage[hang]{footmisc}
\fancypagestyle{plain}{%
  \fancyhf{}
  \fancyhead[R]{\small {\it \jname}, x(x):\thepage--\pageref{LastPage} (\myyear\today)}
  \fancyfoot[R]{\small\bf\thepage }
  \fancyfoot[L]{\fottitle}
  }
\renewcommand{\dmyy}{18}
\renewcommand{\myyear}{2018}
\renewcommand{\today}{}

\begin{document}

\volume{Volume x, Issue x, \myyear\today}
\title{A detailed numerical investigation of two-phase flows inside a planar flow-blurring atomizer}
\titlehead{Numerical investigation of two-phase flows in flow-blurring atomizer}
\authorhead{Y. Ling \& L. Jiang}
\corrauthor[1]{Yue Ling}
\author[2]{Lulin Jiang}
\corremail{stanley\_ling@sc.edu}
\corraddress{Department of Mechanical Engineering, University of South Carolina, Columbia, SC 29208 USA}
\address[1]{Department of Mechanical Engineering, University of South Carolina, Columbia, SC 29208 USA}
\address[2]{Department of Mechanical Engineering, Baylor University, Waco, TX 76798 USA }

\dataO{mm/dd/yyyy}
\dataF{mm/dd/yyyy}

\abstract{
Flow-blurring atomization is an innovative twin-fluid atomization approach that has demonstrated superior effectiveness in producing fine sprays compared to traditional airblast atomization methods. In flow-blurring atomizers, the high-speed gas flow is directed perpendicular to the liquid jet. Under specific geometric and physical conditions, the gas penetrates back into the liquid nozzle, resulting in a highly unsteady bubbly two-phase mixing zone. Despite the remarkable atomization performance of flow-blurring atomizers, the underlying dynamics of the two-phase flows and breakup mechanisms within the liquid nozzle remain poorly understood, primarily due to the challenges in experimental measurements of flow details. In this study, detailed interface-resolved numerical simulations are conducted to investigate the two-phase flows generated by a planar flow-blurring atomizer. By varying key dimensionless parameters, including the dynamic-pressure ratio, density ratio, and Weber number, over wide ranges, we aim to comprehensively characterize their effects on the two-phase flow regimes and breakup dynamics.
}


\maketitle

\section{Introduction} 
Two-phase/twin-fluid atomizers are generally more effective than single-fluid atomizers at the same injection pressure and thus are widely used in spray applications, such as gas turbine engines \citep{Lefebvre_1980a}. One of the most commonly used twin-fluid atomizers is the airblast atomizer, for which the high-speed gas is injected parallel to the liquid jet, and the large velocity difference between the gas and liquid destabilizes and disintegrates the liquid jet into fine droplets. The shear instability is triggered at the gas-liquid interface, which forms interfacial waves that grow spatially \citep{Lasheras_2000a, Marmottant_2004a}. Liquid droplets are eventually formed from the breakup of interfacial waves or the disintegration of the distorted liquid jet  \citep{Ling_2017a, Agbaglah_2017a}. Although airblast atomizers are commonly used in gas turbine engines, they have limitations. The main drawback is that the primary breakup of the liquid jet occurs outside of the nozzle and relies on the spatial development of shear instability. As a result, for high-viscosity liquids, the length of complete atomization is long, and the atomization efficiency is relatively low \citep{Qavi_2021f}.

To overcome the limitations of airblast atomizers, the flow-blurring atomizer was introduced \citep{Ganan-Calvo_2005a}. The fundamental difference between flow-blurring and airblast atomizers lies in the direction of the atomization gas. In airblast atomizers, the gas flow is generally parallel to the liquid jet, while in flow-blurring atomizers, the gas flow is perpendicular to the liquid jet. Under proper operating conditions, the injected gas impinges on the liquid jet and flows back into the liquid nozzle. The gas backflow interacts with the incoming liquid, creating a two-phase bubbly mixing zone inside the nozzle \citep{Agrawal_2013a, Murugan_2021f}. As this two-phase mixture passes the nozzle exit and interacts with the atomizing gas jet, the liquid sheets around the bubbles break and fine droplets are immediately formed just outside the nozzle exit \citep{Jiang_2015a, Jiang_2015b}. Under similar injection conditions and fluid properties, flow-blurring atomizers have demonstrated significantly higher atomization performance compared to conventional airblast atomizers, particularly for highly viscous liquids \citep{Ganan-Calvo_2005a, Simmons_2010h, Sallevelt_2015i, Jiang_2015b, Fisher_2018a, Qavi_2021f}.

It is important to note that simply changing the direction of the gas inlet in the flow-blurring atomizer does not guarantee the flow-blurring atomization mode or the formation of a characteristic backflow region in the liquid nozzle. The interaction between the impinging gas streams and the liquid jet can vary depending on the geometric and flow parameters, leading to a wide variety of breakup modes \citep{Ganan-Calvo_2005a, Khan_2019a, Jaber_2020a}. The transition between different modes depends on parameters, such as the ratio between the gas gap and liquid jet diameter, the gas-to-liquid mass flux ratio, and the Weber and Ohnesorge numbers.

Although the flow-blurring atomizer and its variations \citep{Danh_2019a} show promise in effectively generating fine sprays, the underlying mechanisms are still not fully understood. This is primarily due to the highly unsteady nature of the two-phase flows inside the nozzle, which makes it challenging to obtain detailed flow information through experimental diagnostics \citep{Agrawal_2013a, Murugan_2020j, Murugan_2021f}. Therefore, numerical simulations serve as an important alternative to investigate the flow physics and breakup dynamics inside the nozzle, which are currently unclear.

Several recent attempts have been made to simulate the two-phase flows involved in flow-blurring atomization \citep{Zhao_2019a, Murugan_2020j, Ates_2021b}. However, these simulations failed to capture the characteristic features of flow-blurring atomization observed in experiments, such as the presence of an internal two-phase bubbly mixing zone inside the liquid nozzle. 
{Snapshots of the 3D simulation results \cite{Murugan_2020j} showed gas bubbles in the liquid nozzle, but the formation process of the bubbly mixing zone was not shown. Furthermore, similar to other previous simulations, their results exhibit long, unbroken liquid jets outside the nozzle exit, similar to airblast atomization, which are not consistent with experimental observations.} 
In a recent simulation by \cite{Ates_2021b}, the authors had to adjust the gas injection angle against the liquid flow to reproduce the internal two-phase mixing zone. However, such geometric adjustments are not required in experiments to achieve the flow-blurring atomization mode, raising further questions about the accuracy of these simulations.

The goal of this study is to conduct a detailed numerical investigation of the two-phase flows induced by a planar flow-blurring atomizer under different operating conditions and fluid properties. To accurately capture the sharp gas-liquid interfaces, a geometric Volume-of-fluid (VOF) method is employed. 
{For simulations using VOF methods, numerical breakup will occur when the mesh resolution is insufficient to resolve thin liquid sheets or filaments. Therefore, even though the van der Waals force is not included in the governing equations, primary breakup can be captured. Assuming sufficient mesh resolution, the breakup of filaments and liquid sheets can be well captured by VOF simulations, and good agreement with experiments has been observed \cite{Zhang_2019b, Ling_2017a, Tang_2023a, Ling_2023a}.
}Two-dimensional (2D) simulations are performed for the two-phase flows inside the planar flow-blurring atomizer. 
{
Though the 2D simulations will not fully capture the 3D features, such as the droplet formation and statistics, they are sufficient to capture the interfacial dynamics and instability, which in turn determines the internal-flow pattern, as evident in previous numerical studies for two-phase mixing layer in gas-assisted atomization \citep{Fuster_2013a, Bozonnet_2022a}. 
}
Taking advantage of the relatively low computational cost of {2D simulations}, a large number of simulation cases are carried out to identify the important dimensionless parameters that characterize the two-phase flow regimes and breakup modes. 
The governing equations and numerical methods used in this study are introduced in Section \ref{sec:methods}. In Section \ref{sec:setup}, the simulation setup details, including the computational domain, boundary conditions, and key parameters, are described. The results obtained from the simulations are presented and discussed in Section \ref{sec:results}. Finally, the key findings of the present study are summarized in Section \ref{sec:conc}.

\section{Governing equations and numerical methods}
\label{sec:methods}
The one-fluid approach is employed to resolve the gas-liquid two-phase flows inside the flow-blurring atomizer. The liquid and the gas are treated as one fluid with material properties that change abruptly across the gas-liquid interface. The Navier-Stokes equations for incompressible flow with surface tension are given as
\begin{equation}
  \rho (\delta_t \mathbf{u} + \mathbf{u} \cdot \nabla \mathbf{u}) = -\nabla p + \nabla \cdot (2 \mu \mathbf{D}) + \sigma \kappa \delta_s \mathbf{n},
  \label{eq:NS1}
\end{equation}
\begin{equation}
  \nabla \cdot \mathbf{u} = 0,
  \label{eq:NS2}
\end{equation}
 where $\rho$, $\mu$, $\mathbf{u}$, and $p$ represent density, viscosity, velocity, and pressure, respectively. The strain-rate tensor is denoted by $\mathbf{D}$. The third term of the right-hand side of Eq.\ \eqref{eq:NS1} is a singular term, with a Dirac distribution function $\delta_s$ localized on the interface, and it represents the surface tension. The surface tension coefficient is $\sigma$, and $\kappa$ and $\mathbf{n}$ are the local curvature and unit normal of the interface, respectively.

The volume fraction of liquid $C$ is introduced to distinguish the different phases; $C = 0$ and 1 for cells containing only gas and liquid phases, respectively. The time evolution of $C$ satisfies the advection equation
\begin{equation}
  \delta_t C + \mathbf{u} \cdot \nabla C = 0.
  \label{eq:C1}
\end{equation}

The fluid density and viscosity are determined by 
\begin{align}
  \rho & = C \rho_l + (1 - C) \rho_g\,,
  \label{eq:density1}\\
  \mu &= C \mu_l + (1 - C) \mu_g\, ,
  \label{eq:viscosity1}
\end{align}
where the subscripts $g$ and $l$ correspond to the gas phase and the liquid phase, respectively.

The Navier-Stokes equations (Eqs.\ \eqref{eq:NS1} and \eqref{eq:NS2}) are solved by the open-source multiphase flow solver, \textit{Gerris} \citep{Popinet_2003a, Popinet_2009a}. The \textit{Gerris} solver uses a finite-volume approach based on a projection method. A staggered-in-time discretization of the volume-fraction/density and pressure leads to a formally second-order-accurate time discretization. The advection equation for liquid volume fraction (Eq.\ \eqref{eq:C1}) is solved using a geometric Volume-of-Fluid (VOF) method \citep{Scardovelli_1999a, Popinet_2009a}.  A {quadtree} spatial discretization is used, which gives a very important flexibility by allowing adaptive mesh refinement in user-defined regions. The height-function (HF) method is used to calculate the local interface curvature \citep{Popinet_2009a} and to specify the contact angle on the surface \citep{Afkhami_2009a}. Finally, a balanced-force surface tension discretization is used \citep{Francois_2006a,Popinet_2009a}. Validation studies for the numerical methods and the  solver in solving a wide variety of interfacial multiphase flows can be found in previous studies \citep{Popinet_2009a}.

\section{Simulation setup}
\label{sec:setup}
\begin{figure}[tbp]
 \centering{\includegraphics[trim=1.2in 2in 1.2in 2.in,clip, width=0.95\textwidth]{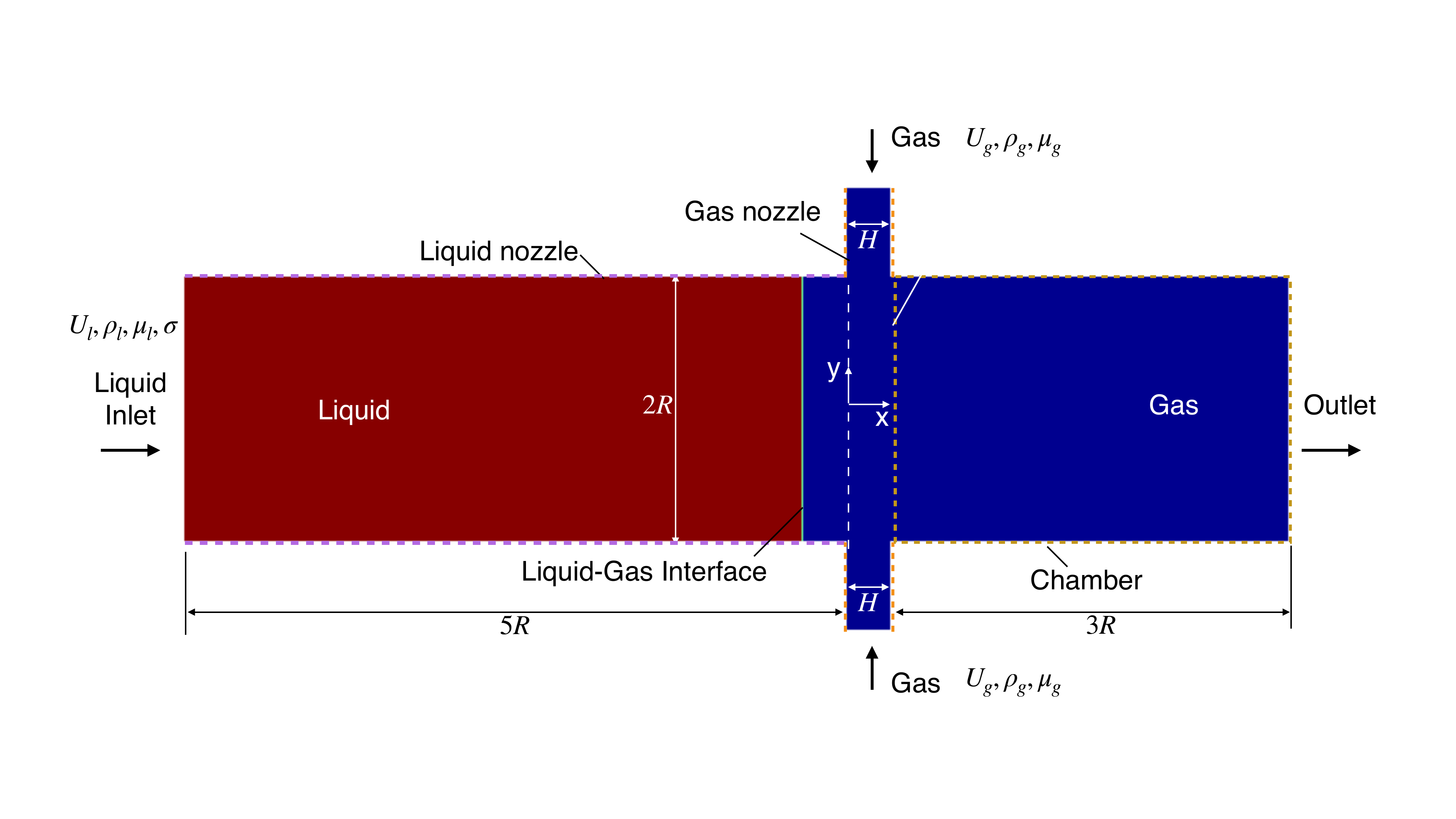}}
 \caption{Simulation setup.}
 \label{fig:setup}
\end{figure}

The 2D computational domain is depicted in Fig.~\ref{fig:setup}. The flow-blurring atomizer (injector) consists of one liquid nozzle and two gas nozzles. The liquid enters the liquid nozzle from the left boundary with a uniform velocity $U_l$, while the the gases enter the top and bottom nozzles with identical {uniform} velocities $U_g$. The widths of the liquid and gas inlets are $2R$ and $H$, respectively. The nozzle walls are treated as no-slip surfaces with the contact angle specified. The injector exit is connected to a chamber which is initially filled with stationary gas and the right surface of the domain serves as the outlet, with the pressure set to zero. 
{
To focus the computational resources on the internal flow, a small gas chamber has been used. The chamber height is 6\(H\) and the length is \(9H\), which are both significantly larger than the flow length scale set by the gas nozzle width \(H\). The current chamber size is sufficient to resolve the flow inside the nozzle and near the nozzle exit. To capture the turbulent dispersion of droplets further downstream of the nozzle exit, a larger chamber size will be required, which will be relegated to future work. 
}

The origin of the coordinate is located at the center of the liquid nozzle exit, as shown in Fig.~\ref{fig:setup}. The colors in indicates the initial distribution of the liquid volume fraction $C$, with red and blue corresponding to liquid and gas, respectively. The liquid-gas interface is initially flat and a distance of $H$ away from liquid nozzle exit, namely $x=-H$. The liquid and gas densities are denoted as $\rho_l$ and $\rho_g$, respectively, and the viscosities are $\mu_l$ and $\mu_g$, respectively. The surface tension is $\sigma$. The contact angle on the nozzle wall is general set as $\theta=70$\textdegree, and a sensitivity analysis of $\theta$ is performed. The fluid properties, inlet velocities, and geometric parameters are listed in Table \ref{tab:phys_para}.  

\begin{table}[tbp]
\begin{center}
\begin{tabular}{|cccccccccc|} \hline
$\rho_l$ & $\rho_g$  & $\mu_l$ & $\mu_g$ & $\sigma$ & $U_l$ & $U_g$ & $H$ & $R$ & $\theta$\\ 
kg/m$^3$ & kg/m$^3$ & Pa\ s & Pa\ s & N/m & m/s &m/s & mm & & \textdegree  \\ \hline
1000 & 3.125 - 50 & $10^{-3}$ & $5\times 10^{-3}$ & 0.05 & 0.0625 - 2	& 40 &0.05 - 3.2 & 3$H$ & 30 - 90\\
\hline
\end{tabular}
\end{center}
\caption{Physical parameters used in the present simulations. }
\label{tab:phys_para}
\end{table}

The key dimensionless parameters that determine the characteristics of two-phase flows produced by the flow-blurring atomizer are the density ratio ($\mathrm{r}=\rho_l/\rho_g$), viscosity ratio ($\mathrm{m}=\mu_l/\mu_g$), gas-to-liquid inlet size ratio ($\eta=H/R$), dynamic-pressure ratio ($M=(\rho_g U_g^2)/(\rho_l U_l^2)$), gas Reynolds number ($\mathrm{Re}=\rho_g U_g H/\mu_g$), and Weber number ($\mathrm{We}=\rho_g U_g^2H/\sigma$). Another commonly used parameter is the gas-to-liquid mass inflow ratio ($\mathrm{GLR}=\dot{m}_g/\dot{m}_l$), which can be calculated from the dimensionless parameters defined above as $\mathrm{GLR}=\eta \sqrt{M/r}$.

In the present study, the values of $\rho_l$, $\mu_l$, $\mu_g$, and $\sigma$ are fixed, while $U_l$, $H$, and $\rho_g$ are varied for the parametric studies of $M$, $\mathrm{We}$, and $\mathrm{r}$. When $H$ is varied, $R$ is correspondingly adjusted to maintain a constant $\eta=1/3$. Furthermore, $\mathrm{m}=0.2$ is kept constant throughout the study. The ranges of dimensionless parameters considered are listed in Table \ref{tab:dmls_para}.

\begin{table}[tbp]
\begin{center}
\begin{tabular}{|cccccc|} \hline
$M$ & $\mathrm{We}$  & $\mathrm{r}$ & $\mathrm{m}$ & $\mathrm{Re}$ & $\eta$\\  \hline
5 - 5120 & 20 - 1280 & 20 - 320 & 20 & 500 - 32,000 & 1/3 \\
\hline
\end{tabular}
\end{center}
\caption{Physical parameters used in the present simulations. }
\label{tab:dmls_para}
\end{table}

The quadtree mesh is utilized to discretize the computational domain, and the maximum refinement level determines the finest mesh allowed. The minimum cell size $\Delta$ used here is equivalent to $H/128$. The dynamic refinement or coarsening of the mesh relies on the gradient of $C$ and the estimated discretization errors of all velocity components. Grid refinement studies have been performed by varying $H/\Delta=64$ to 256, see \ref{sec:grid_refinement}. The results show that the mesh resolution $\Delta = H/128$ is sufficient to resolve the two-phase flows for the ranges of parameters considered. 

{
The \(C\) field is used to identify and tag connected liquid cells. Cells with \(C>0\) that are connected will be assigned the same tag \citep{Herrmann_2010a}, from which the volume (or area for 2D simulations) of individual liquid structures can be determined.
}
In order to {improve} the numerical stability of the simulations, {any liquid fragments} with a size smaller than approximately 4 cells are removed, which is a common practice in Volume-of-Fluid (VOF) simulations \citep{Shinjo_2010a, Ling_2017a}. {For the present 2D simulations, ``droplets" formed in the atomization process are actually cylinders and do not truly represent physical droplets. Therefore, statistical data regarding droplet sizes are not collected here. A comprehensive investigation of the droplet size statistics is deferred to future work.}

\section{Results}
\label{sec:results}
\subsection{Characterization of two-phase flow regimes}
\begin{figure}[tbp]
 \centering{\includegraphics[trim=.0in 0in 0in 0in,clip, width=0.99\textwidth]{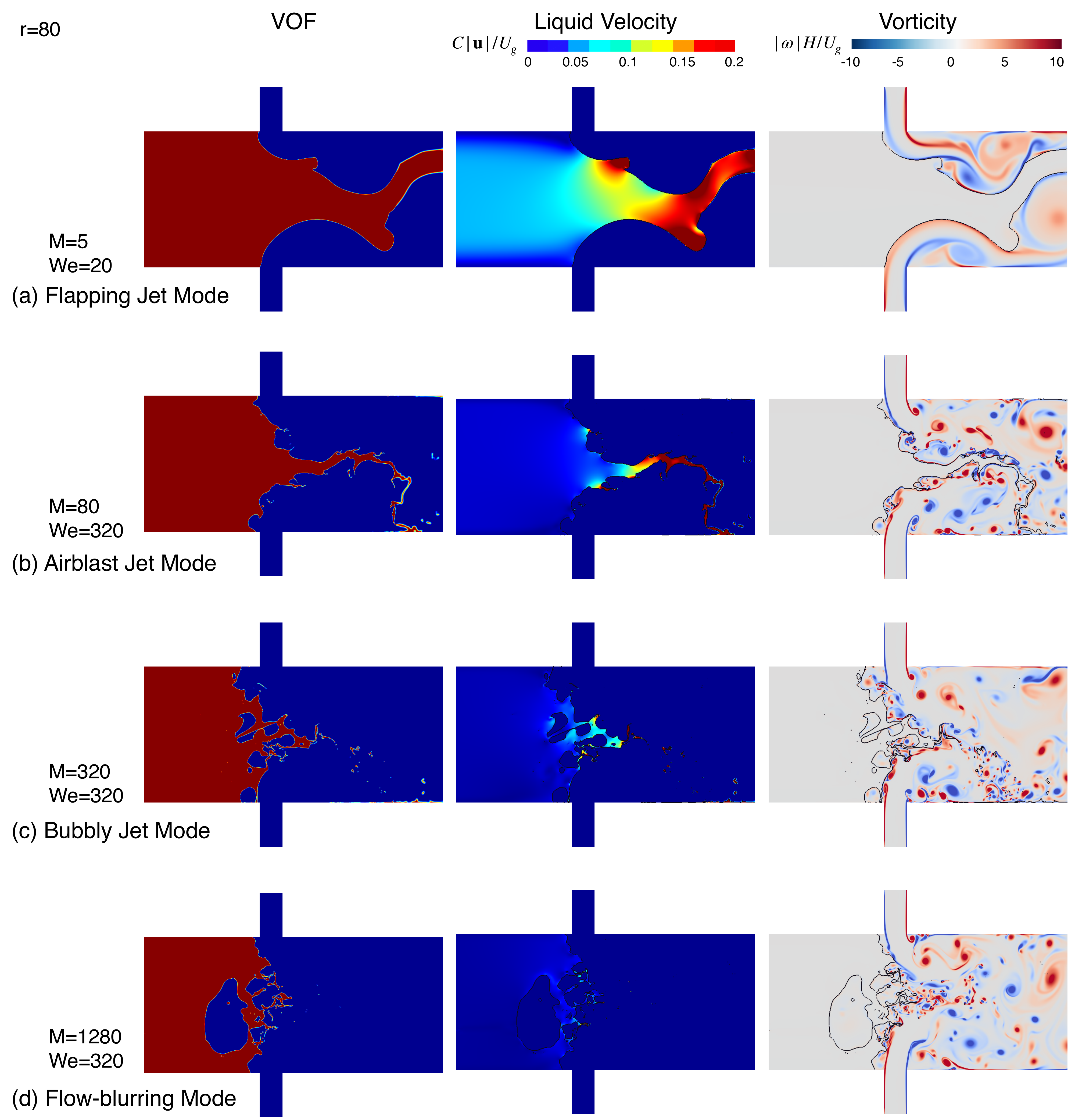}}
 \caption{Representative snapshots of the VOF function ($C$), liquid velocity magnitude ($c|\mathbf{u}|$), and vorticity ($\omega$) for different $M$ and $\mathrm{We}$, which can be used identified different breakup modes. The density ratio is $\mathrm{r}=80$ for all cases shown. }
 \label{fig:breakup_mode}
\end{figure}

Different breakup modes/regimes can be clearly identified from the simulation results for different $M$ and $\mathrm{We}$. Representative snapshots of VOF ($C$), liquid velocity magnitude ($C|\mathbf{u}|$), and vorticity ($\omega$) are shown in Fig.~\ref{fig:breakup_mode} to depict the two-phase flows inside the flow-blurring injector and in the near field outside of the atomzier. 

\begin{itemize}
\item \emph{Flapping-jet regime} When both $M$ and $\mathrm{We}$ are low, the interaction between the gas and liquid jets are not strong enough to disintegrate the liquid jet. As a result, a long unbroken liquid jet is formed outside the injector, see Fig.~\ref{fig:breakup_mode}(a). As the liquid jet is squeezed by the two gas jets, and the thickness of the liquid jet decreases along $x$ axis. Due to mass conservation, the liquid velocity increases along $x$ as well. This flow pattern is also referred to as the flow-focusing mode in previous studies \citep{Ganan-Calvo_2005a}. The liquid jet flaps due to the unstable sinuous mode \citep{Deshpande_2015o, Delon_2018a}. Since a small chamber is used there, so the subsequent breakup of the planar liquid jet is not resolved. It is expected that the liquid sheet will break as the amplitude of sinuous perturbation grows to be larger than the sheet thickness, producing filaments and droplets \citep{Deshpande_2015o}. Though shear instability and the resulting interfacial waves are observed on the jet surface, the interfacial waves rarely break. {From the vorticity field, it is evident that the gas shear layers near the interfaces are generally stable. Flow separation and vortices are induced only on the crest of the interfacial waves.}

\item \emph{Airblast-jet regime:} As \(M\) and \(\mathrm{We}\) increase, the breakup mode transitions to a different regime, as shown in Fig.~\ref{fig:breakup_mode}(b). Firstly, the length of the unbroken jet significantly reduces as \(M\) increases from 5 to 80. {Additionally, due to the increase in \(\mathrm{We}\), surface tension becomes less effective in resisting the development of shear instability on the liquid jet surface. This interfacial instability leads to interfacial waves, which break as they are advected downstream. Simultaneously, the vorticity field shows that the gas stream becomes more unstable due to perturbations from the wavy interfaces.} Consequently, the jet morphology closely resembles that observed in airblast atomization of a liquid jet with parallel coflowing gas streams \citep{Matas_2011a, Fuster_2013a, Ling_2017a, Ling_2019a}, despite the gas inflow being perpendicular to the liquid jet in the present flow-blurring injector. At high \(\mathrm{We}\), the influence of surface tension on the shear interfacial instability becomes secondary, and the length of the unbroken liquid jet primarily depends on the dynamic pressure continuity on the jet surfaces \citep{Lasheras_1998a}. Thus, the length of the unbroken jet decreases as \(M\) increases. In conventional coaxial airblast atomization of cylindrical liquid jets, the length of the unbroken jet scales with \(M^{-1/2}\) \citep{Lasheras_1998a}. It would be intriguing for future studies to determine whether this scaling relation remains valid for the flow-blurring atomizer in the airblast-jet regime, though it is beyond the scope of the current study. While the flapping of the small unbroken liquid jet also contributes to droplet formation, the dominant mechanism for droplet formation in this regime is the breakup of interfacial waves on the jet surfaces.

\item \emph {Bubbly-jet regime} When \(M\) is further increased to 320, while \(\mathrm{We}\) remains the same as in Fig.~\ref{fig:breakup_mode}(b), a distinct breakup mode emerges, see Fig.~\ref{fig:breakup_mode}(c). The length of the unbroken liquid jet is further reduced. More importantly, gas bubbles start to appear in the liquid jet near the liquid nozzle exit. The shear-induced interfacial waves on the jet surfaces grow faster with increasing \(M\), and the rolling interfacial waves entrain bubbles into the liquid jet, similar to the bubble entrainment in breaking waves \citep{Deike_2016a}. As the bubbly jet interacts with the gas jets, the bubbles inside become distorted and eventually burst. 
{The interaction between the gas flows and the bubbly jet is better shown in the vorticity plot. A small amount of atomizing gas is also observed flowing back into the liquid nozzle near the wall.}
The rupture of thin liquid sheets between different bubbles or between a bubble and the jet surface contributes to the formation of droplets. Since van der Waals forces are not considered in the present simulations, the minimum cell size acts as the numerical cutoff length scale to artificially pinch the two interfaces of a liquid sheet when its thickness becomes comparable to the cell size. The rupture of a stationary liquid sheet due to the interaction between capillary and van der Waals forces does not occur until the sheet thickness decreases to about \(10^2\) nm \citep{Oron_1997o}, whereas the rupture thickness for a dynamic liquid sheet has been observed to be significantly larger, around \(O(1)\) \textmu m \citep{Opfer_2014a, Neel_2018r, Jackiw_2022a}. The breakup of bulk liquid in this regime is similar to, but not identical to, the flow-blurring mode because even though bubbles are entrained in the bulk liquid, an unbroken jet still exists outside of the nozzle.

\item \emph{Flow-blurring regime:} Finally, when \(M\) exceeds a certain threshold, see Fig.~\ref{fig:breakup_mode}(d), the full flow-blurring mode is attained. This regime is characterized by the presence of a bubbly two-phase mixing zone inside the liquid nozzle. The flow-blurring mode is accurately reproduced with the gas inflow normal to the liquid jet, without requiring adjustments to the gas inflow direction as seen in other numerical studies \citep{Ates_2021b}. As the bubbly mixture interacts with the gas jets, the bubbles become distorted, and thin liquid sheets between them rupture, resulting in the formation of fine droplets, similar to the bubbly-jet regime. The key distinction is that there is no unbroken liquid jet present outside the injector exit. {Furthermore, from the liquid velocity fields (the central column of Fig.~\ref{fig:breakup_mode}), it can be seen that, in the other three regimes, unbroken bulk liquids exist outside the liquid nozzle. These bulk liquids are accelerated by the fast gas flows and thus display high velocity magnitudes. However, the bulk liquid breakup is complete at the nozzle exit in the flow-blurring mode, as a result, structures with high liquid velocity are not observed. 
The highly unsteady and chaotic two-phase flows near the liquid nozzle exit can be clearly seen from the vorticity field, which are responsible for the phenomenon of flow "blurring" observed from outside the nozzle in experimental observations.}

\end{itemize}

\subsection{Effect of dynamic pressure ratio on the transition to flow-blurring regime}

\begin{figure}[tbp]
 \centering{\includegraphics[trim=3.3in 0.5in 2.2in 0.5in,clip, width=0.99\textwidth]{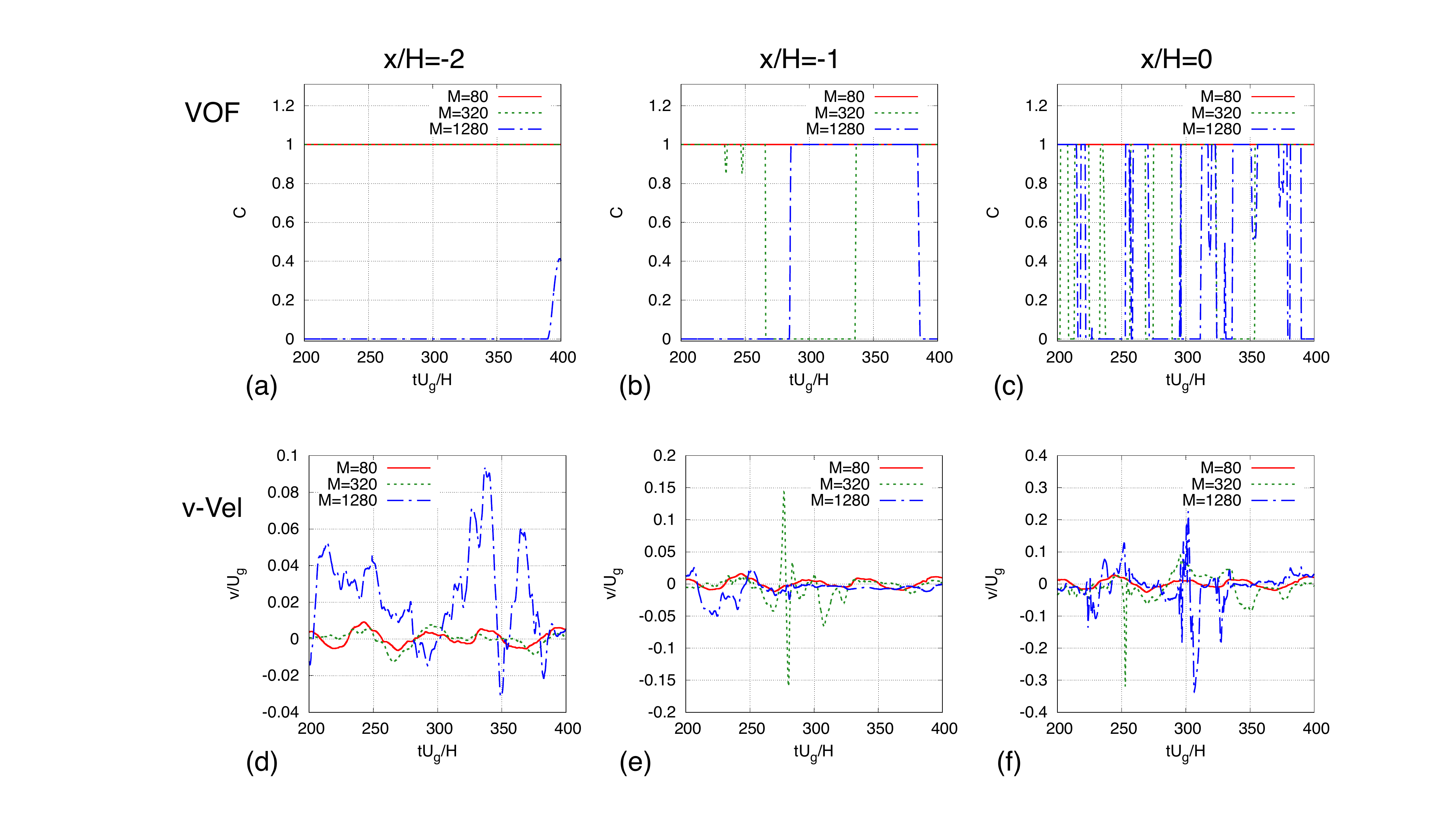}}
 \caption{Temporal evolutions of (a)-(c) the VOF function ($C$) and (d)-(f) $v$-Velocity at different locations inside the liquid nozzle (a,d) $x/H=-2$, (b,e) $x/H=-1$, and (c,f) $x/H=0$ on the central axis ($y=0$).  }
 \label{fig:back_flow}
\end{figure}

The results shown in Figs.~\ref{fig:breakup_mode}(b)-(d) correspond to different $M$, while $\mathrm{We}=320$ and $\mathrm{r}=80$ remain fixed. These results clearly demonstrate the significant influence of the parameter $M$ on the transition of the breakup mode from the airblast-jet regime to the flow-blurring regime. To gain a better understanding of the flow characteristics within the two-phase mixing zone, the temporal evolution of the VOF function ($C$) and the $y$-component of velocity ($v$) at three different locations ($x/H=0$, $-1$, and $-2$) along the $x$-axis ($y=0$) inside the liquid nozzle are presented in Fig.~\ref{fig:back_flow}. It can be observed that $C=1$ throughout the entire duration at all three locations for $M=80$, indicating that no gas penetrates back into the liquid nozzle along $x$-axis. Consequently, the velocity $v$ at these locations solely represents the liquid velocity. {The value of \(v\) oscillates over time about zero, as the mean flow is symmetric with respect to the \(x\)-axis.} The dominant oscillation mode exhibits a period of approximately $50 H/U_g$, which is determined by the dominant mode of the shear instability outside the nozzle. Similar to airblast atomization for parallel gas and liquid streams, for high $M$ the shear instability is absolute, allowing the perturbations to propagate back into the liquid inside the nozzle \citep{Matas_2011a, Fuster_2013a, Ling_2019a, Jiang_2021a}.

\begin{figure}
 \centering{\includegraphics[trim=0.in 0.in 0.in 0.in,clip, width=0.99\textwidth]{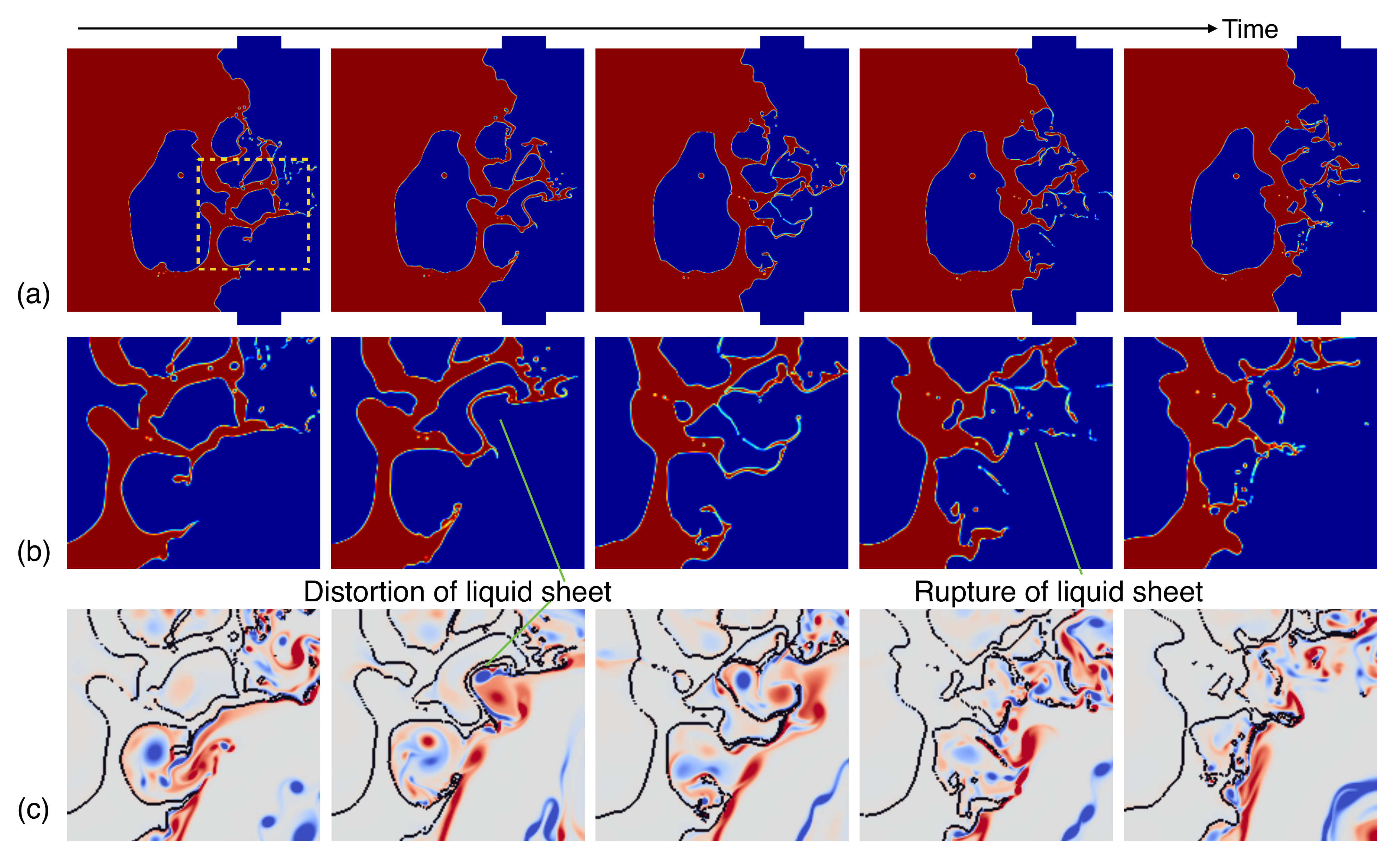}}
 \caption{Time snapshots of (a) the VOF function ($C$) near the liquid nozzle exit, and closeups of (b) $C$ and (c) vorticity ($\omega$) to depict the droplet formation due to bubble bursting. The color scale for the vorticity is the same as Fig.~\ref{fig:breakup_mode}. }
 \label{fig:bubbles_bursting}
\end{figure}

As $M$ increases to 320 and 1280, it is shown that the value of $C$ at the liquid nozzle exit ($x/H=0$) oscillates between 0 and 1. These oscillations are due to the intense breakup of liquid sheets near the nozzle exit and the resulting highly unsteady flows. It can be observed from Fig.\ref{fig:back_flow}(a) that $C$ at $x/H=-2$ is predominantly 0 for the $M=1280$ case, indicating that this location ($x/H=-2,y=0$) is situated within the large primary bubble inside the liquid nozzle (as also shown in Fig.\ref{fig:bubbles_bursting}). Although there is gas flow inside the large bubble, the velocity magnitude is relatively small compared to the gas velocity near the liquid nozzle exit, as depicted in Figs.~\ref{fig:back_flow}(d) and (f). For the $M=320$ case, it can be observed that $C$ remains equal to 1 at $x/H=-2$, whereas for $M=1280$, $C$ mostly takes on a value of 0. This suggests that the penetration depth of the gas bubbles ($L_b$) into the liquid nozzle generally increases with increasing $M$, and a sufficiently large penetration depth is required to place the flow in the flow-blurring regime. A quantitative measurement of the penetration depth will be provided in the subsequent section.

\begin{figure}[tbp]
 \centering{\includegraphics[trim=1.3in 2.6in 1.9in 2.0in,clip, width=0.99\textwidth]{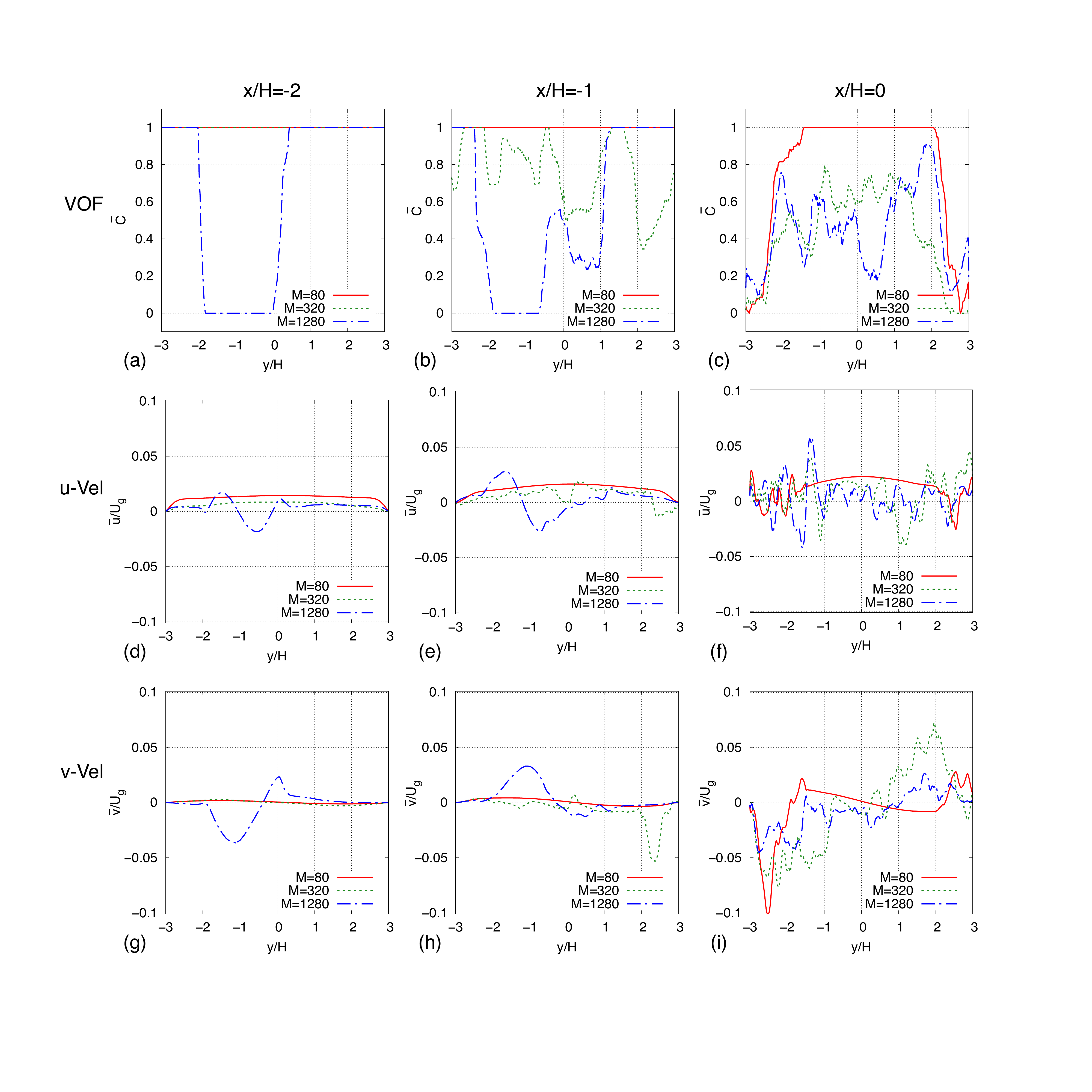}}
 \caption{Spatial variation in $y$ directions for time-averaged properties: (a)-(c) the VOF function ($\overline{C}$), (d)-(f) $\overline{u}$, and (g)-(i) $\overline{v}$ at different $x$ locations inside the liquid nozzle (a,d,g) $x/H=-2$, (b,e,h) $x/H=-1$, and (c,f,i) $x/H=0$.  }
 \label{fig:back_flow_mean}
\end{figure}

{
The spatial variations in the \(y\) direction of the time-averaged VOF function and velocity components, \(\overline{C}\), \(\overline{u}\), and \(\overline{v}\), where \(\overline{(\ )}\) denotes time-averaged quantities, at different \(x\) locations, are shown in Fig.~\ref{fig:back_flow_mean}. At the liquid nozzle exit \(x/H=0\), \(\overline{C}=1\) near the central axis of the nozzle for \(M=80\), which corresponds to the unbroken liquid jet. In contrast, \(\overline{C}\) is generally lower than 1 for the other two cases with large \(M\) due to the existence of the bubbly mixture at the nozzle exit. At locations inside the liquid nozzle, namely \(x/H<0\), \(\overline{C}=1\) for \(M=80\), confirming that no gas penetrates back into the liquid nozzle. However, for \(M=1280\), \(\overline{C}=0\) is observed in certain ranges of \(y\), see for example \(-2<y/H<0\) in Fig.~\ref{fig:back_flow_mean}(a). This observation indicates that this region is always occupied by the primary gas bubble formed in the flow-blurring mode, as seen in Fig.~\ref{fig:breakup_mode}(d). For the intermediate \(M=320\), \(\overline{C}\) lies between 0 and 1, suggesting that while bubbles appear inside the liquid nozzle, they don't consistently occupy specific regions as does the primary bubble in the flow-blurring mode. Instead, these bubbles move outward with the liquid and eventually burst, as shown in Fig.~\ref{fig:breakup_mode}(c).
}

{
The gas backflow and its interaction with the bulk liquid are also depicted by the time-averaged fluid velocities \(\overline{u}\) and \(\overline{v}\). For \(M=80\), \(\overline{u}\) and \(\overline{v}\) are smooth inside the nozzle. It is seen that \(\overline{u}\) is generally positive except in the near-wall regions at \(x/H=0\) where the gas interacts with the liquid jet, as shown in Fig.~\ref{fig:breakup_mode}(b), while \(\overline{v}\) is approximately zero. In contrast, for \(M=1280\), the velocities exhibit large-amplitude low-wavenumber variations, which are induced by the gas's rotational motion inside the primary bubble in the flow-blurring mode, as illustrated in Fig.~\ref{fig:breakup_mode}(d). Velocity fluctuations are also observed for \(M=320\) (see Fig.~\ref{fig:back_flow_mean}), due to the distribution of small bubbles in the bulk liquid.  Yet \(\overline{u}\) generally remains positive, which differs from the flow-blurring case \(M=1280\). For both \(M=320\) and 1280, high wavenumber fluctuations appear in all variables at \(x/H=0\), highlighting the intense breakup of the bubbly mixture.
}

To provide further insight into the breakup of the bubbly mixture near the nozzle exit, sequential closeup snapshots are presented in Fig.~\ref{fig:bubbles_bursting}. These close-ups, focusing on a region near the liquid nozzle exit as highlighted in Fig.~\ref{fig:bubbles_bursting}(a), illustrate the interaction between the bubbly mixture and the two gas jets from the top and bottom. For the centrally positioned bubble depicted in Figs.~\ref{fig:bubbles_bursting}(b) and (c), the liquid sheet on the bottom right undergoes distortion as the bubble interacts with the vortex induced by the bottom gas jet. Consequently, the liquid sheet deforms and curves towards the left, eventually impinging on the bubble's left surface, causing the bubble to split into two smaller child bubbles. Shortly thereafter, the child bubbles burst, resulting in the fragmentation of the surrounding liquid sheets into multiple droplets. The sizes of these droplets are comparable to the thickness of the ruptured liquid sheet. Following the rupture of the liquid sheet, the bubble transforms into a gas cavity. Notably, the high-speed gas enters the cavity and further distorts the unbroken liquid sheets, leading them to undergo complete fragmentation into droplets. {Though the results shown here are obtained by 2D simulations, the breakup mechanisms in 3D will be similar. }

\subsection{Effect of Weber number on the transition to flow-blurring regime}
\begin{figure}
 \centering{\includegraphics[trim=0.in 0.in 0.in 0.in,clip, width=0.99\textwidth]{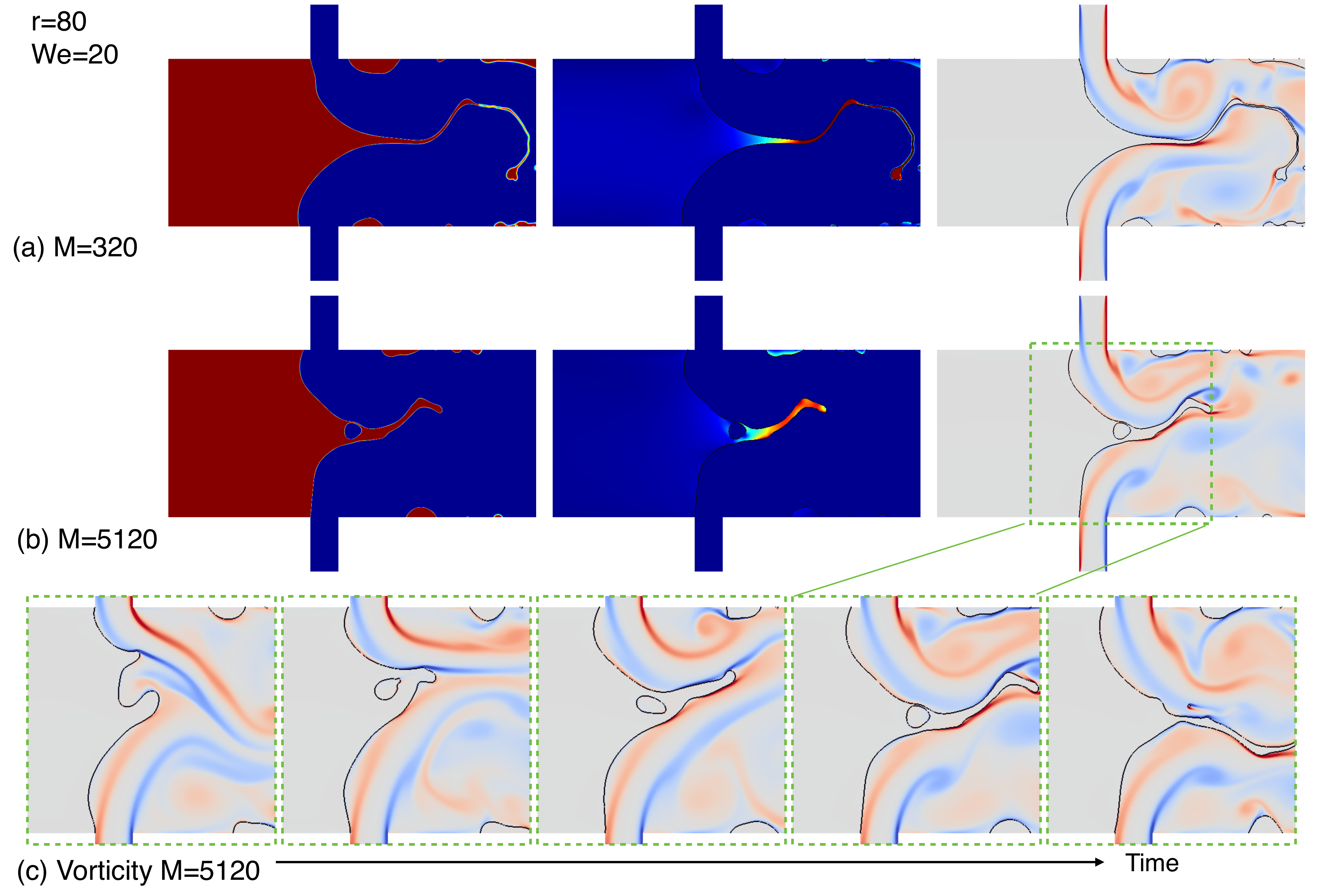}}
 \caption{(a)-(b) Representative snapshots of the VOF function ($C$), liquid velocity magnitude ($c|\mathbf{u}|$), and vorticity ($\omega$) for different $\mathrm{We}$, while $\mathrm{r}=80$ and $M=320$ are fixed. The color scales for the liquid velocity and the vorticity are the same as Fig.~\ref{fig:breakup_mode}. (c) Time evolution of vorticity and interfaces to show the entrainment and the eventual burst of a bubble in the liquid jet. }
 \label{fig:We20}
\end{figure}

Though $M$ is the dominant factor, the influence of $\mathrm{We}$ is also significant, especially for low $\mathrm{We}$ cases. Figure~\ref{fig:We20} shows the results for $M=320$ and $5120$, while maintaining a low Weber number of $\mathrm{We}=20$. As previously shown in Fig.~\ref{fig:breakup_mode}(c), for a high $\mathrm{We}=320$, the case with $M=320$ corresponds to the bubbly-jet regime. However, in this case, with the same $M$ but a significantly lower $\mathrm{We}$, it falls into the flapping-jet regime. Furthermore, even when $M$ is further increased to a large value of $M=5120$, it can be observed that the case with $\mathrm{We}=20$ remains in the flapping-jet regime. Conversely, for $\mathrm{We}=320$, such a high $M$ will place the case in the flow-blurring regime, similar to Fig.~\ref{fig:breakup_mode}(d).

As the Ohnesorge number is relatively low for the considered cases, surface tension becomes the dominant mechanism in resisting interface deformation and liquid breakup. A low $\mathrm{We}$ indicates a stronger suppression of shear-induced interfacial waves. Consequently, the interfacial waves do not develop rapidly enough to roll and entrain bubbles into the liquid. Therefore, even with high gas dynamic pressure, the high-speed gas does not flow back into the liquid nozzle. Even in the case of a very high $M=5120$, the interfacial waves may occasionally entrain a single bubble in the liquid jet, as shown in the sequential snapshots in Fig.~\ref{fig:We20}(c). However, the bubble remains within the liquid jet outside the liquid nozzle exit. The bursting of this bubble generates small droplets, but their contribution to the overall droplet formation is minimal. The majority of droplets are formed due to the flapping liquid jet, similar to other cases in the flapping-jet regime.

\begin{figure}
 \centering{\includegraphics[trim=0.in 0.in 0.in 0.in,clip, width=.99\textwidth]{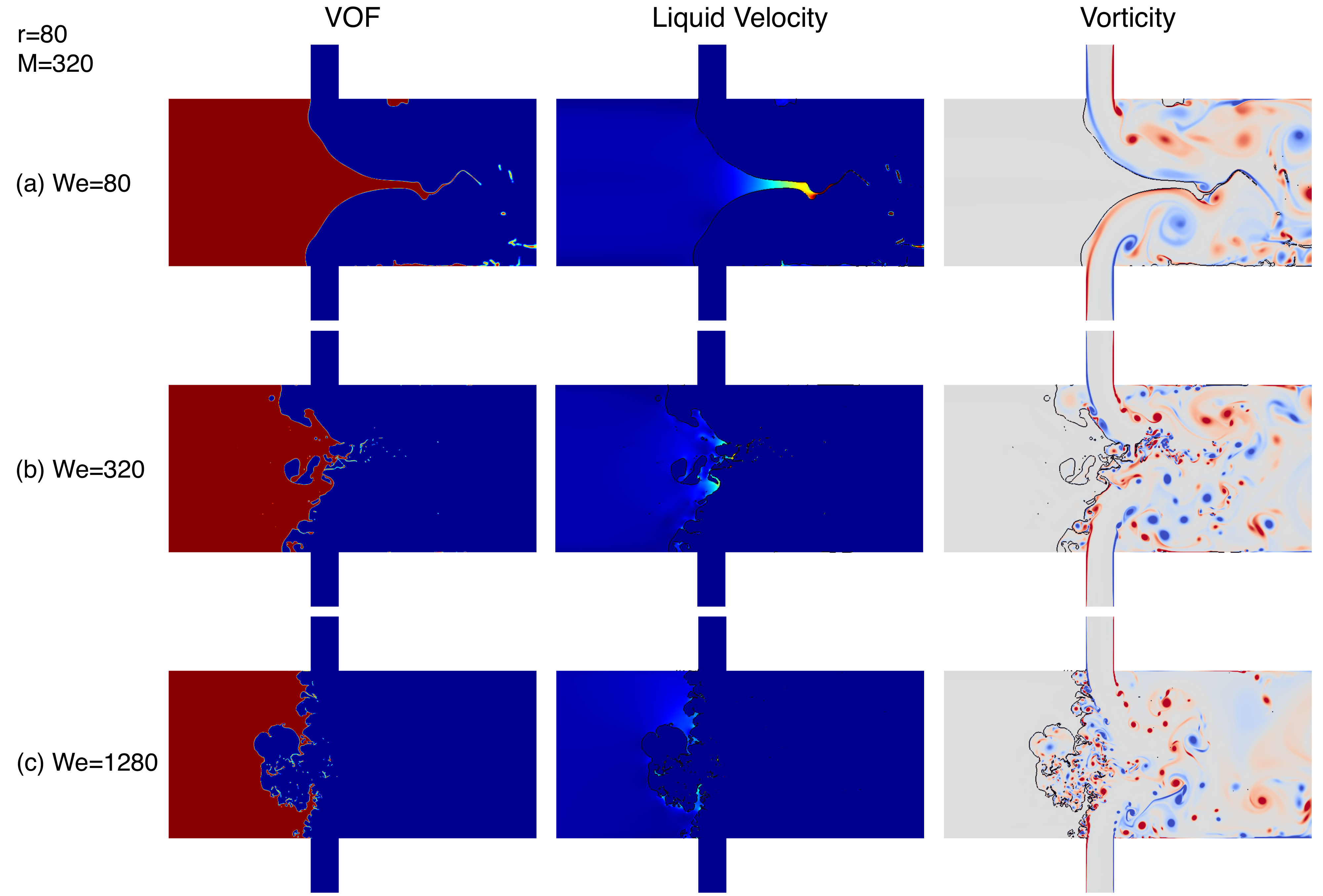}}
 \caption{Representative snapshots of the VOF function ($C$), liquid velocity magnitude ($c|\mathbf{u}|$), and vorticity ($\omega$)  for different $\mathrm{We}$, while $\mathrm{r}=80$ and $M=320$ are fixed. The color scales for the liquid velocity and the vorticity are the same as Fig.~\ref{fig:breakup_mode}. }
 \label{fig:effect_We}
\end{figure}

For cases with high $\mathrm{We}$, the effect of $\mathrm{We}$ is generally less profound. Nevertheless, the present simulation results seem to indicate that the effect of $\mathrm{We}$ is also important for the bubbly-jet regime with moderate $M$. Because the bubbly-jet regime is a transition regime between the more stable airblast-jet and flow-blurring modes, the regime is more sensitive to the change in $\mathrm{We}$. The results for the same $M$ but different $\mathrm{We}$ are shown in Figs.~\ref{fig:effect_We}. It can be clearly seen that the case with $\mathrm{We}=320$ is in the bubbly-jet regime. For a lower $\mathrm{We}$, i.e., $\mathrm{We}=80$, the two-phase flow transitions to the airblast-jet or even flapping-jet regimes. As explained above, the stronger surface tension effect hinders the development of shear-induced interfacial waves, preventing the waves from entraining bubbles into the liquid and forming a bubbly jet. In contrast, when $\mathrm{We}$ is increased to 1280, it is observed that the two-phase flow transitions to the flow-blurring regime. The reduced surface tension effect enhances the breakup of interfacial waves and liquid sheets separating bubbles, making it easier for the gas flow to flow back into the liquid nozzle.

\subsection{Regime map on $M$-$\mathrm{We}$ plane}
\begin{figure}
 \centering{\includegraphics[trim=0.5in 2.in 0.5in 2.in,clip, width=.99\textwidth]{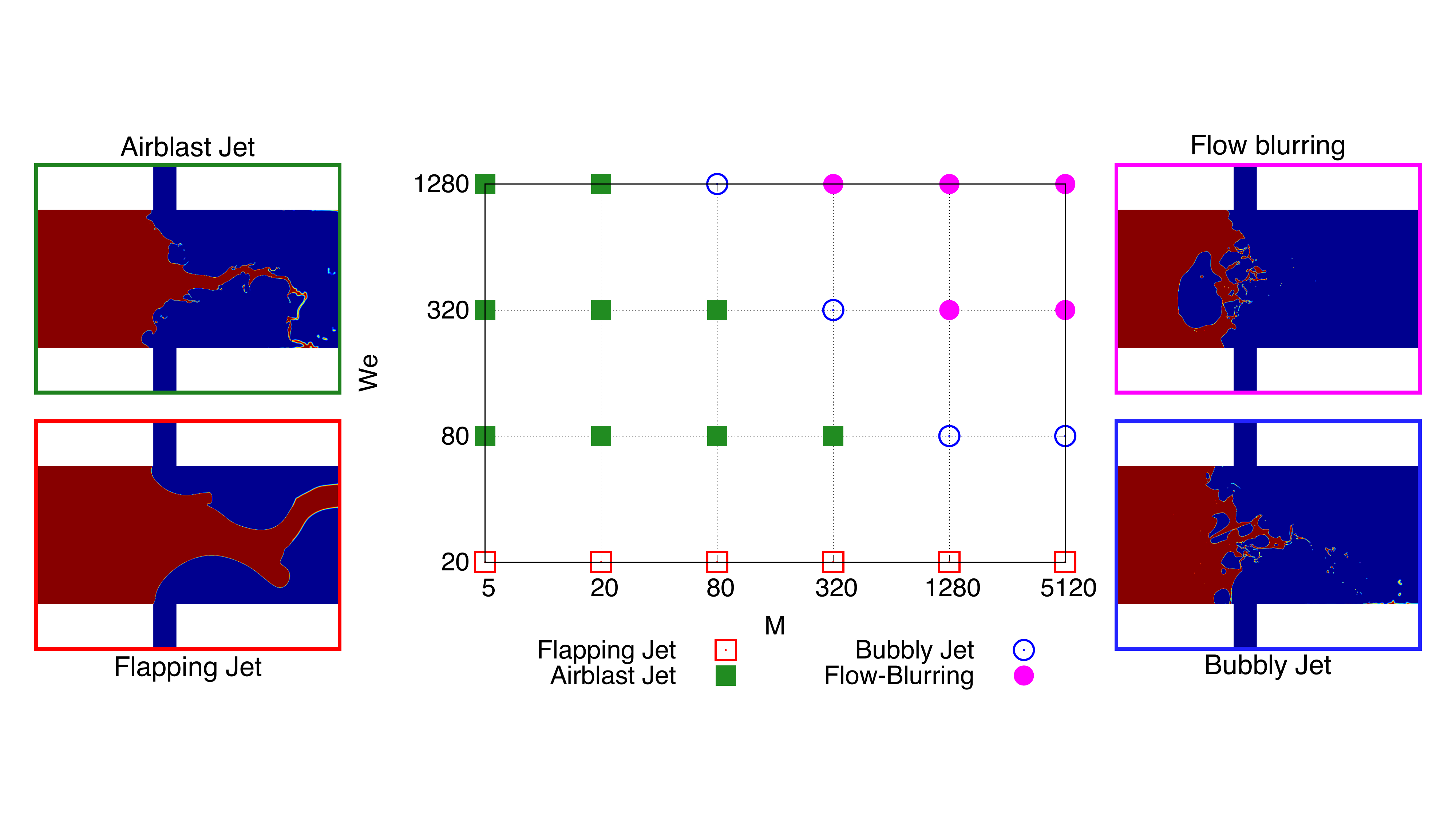}}
 \caption{Different two-phase flow and breakup regimes on the $M$-$\mathrm{We}$ map for $\mathrm{r}=80$. }
 \label{fig:regime_map}
\end{figure}

The results of the parametric simulations for $5<M<5120$ and $20<\mathrm{We}<1280$ are summarized in Fig.~\ref{fig:regime_map}, indicating the different regimes in the parameter space. Here, $\mathrm{r}=80$ is fixed. When $\mathrm{We}$ is very low, the effect of $M$ is negligible, and the two-phase flow near the nozzle exit is always in the flapping regime. For high $M$ and $\mathrm{We}=20$, single or a very small number of bubbles can occasionally form in the liquid, as shown in Fig.~\ref{fig:We20}(b). Nevertheless, the jet mainly breaks up due to flapping instability, and we still consider those cases to be in the flapping-jet regime.

When $\mathrm{We}$ is above 80, it can be seen that $M$ is an important parameter in determining the two-phase flow regimes. For a given $\mathrm{We}$, the flow transitions from the airblast-jet regime to the bubbly-jet regime, and finally to the flow-blurring regime. The region for the bubbly-jet regime is generally narrow since it is a transition between the airblast-jet and flow-blurring regimes. To achieve a fully flow-blurring breakup mode, both $M$ and $\mathrm{We}$ need to be sufficiently large.

\begin{figure}
 \centering{\includegraphics[trim=6.3in 6.5in 6.3in 6.5in,clip, width=.99\textwidth]{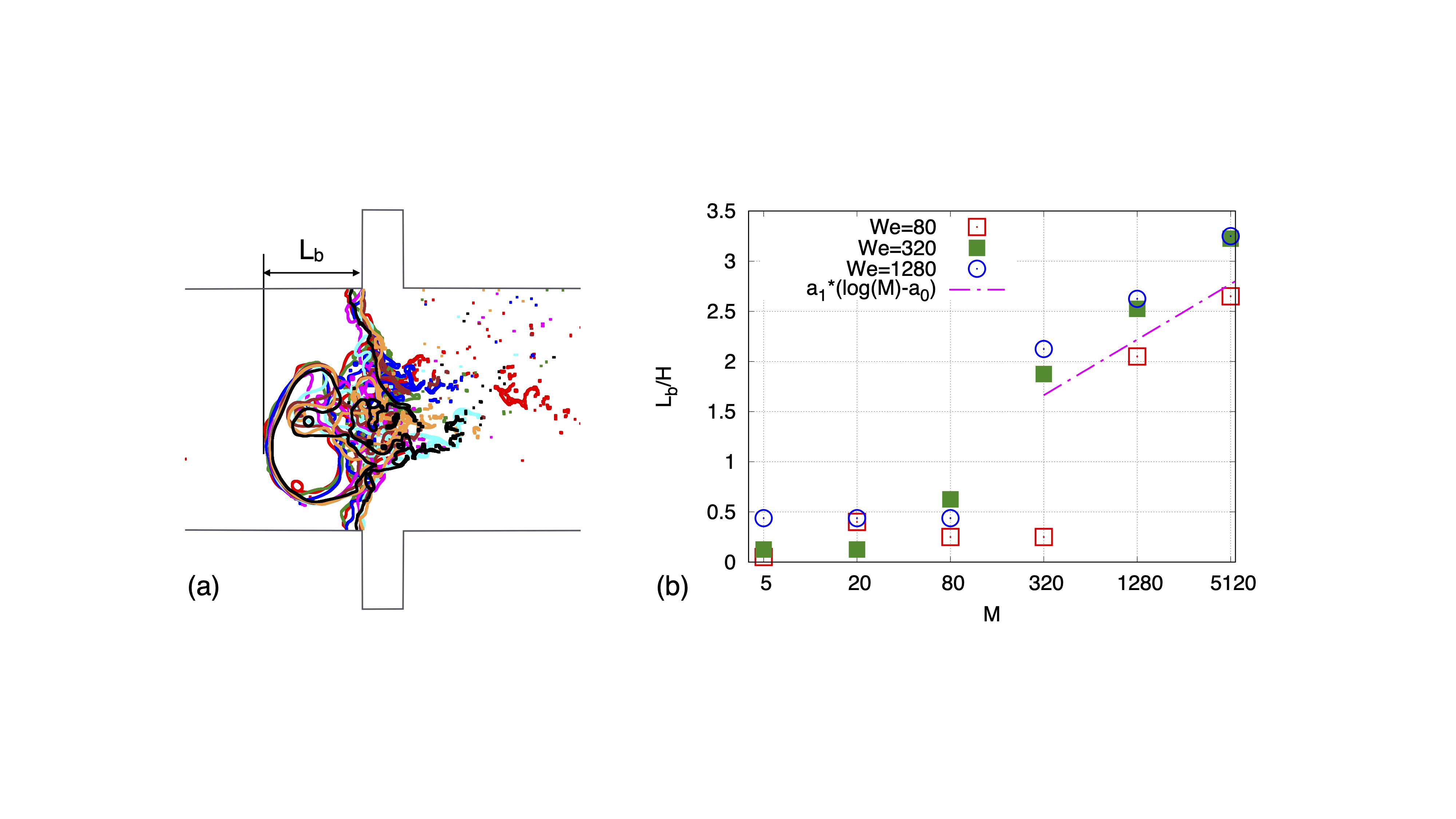}}
 \caption{(a) Gas-liquid interfaces at different times for the case $\mathrm{r}=80$, $M=1280$, $\mathrm{We}=320$ to demonstrate the maximum length of gas bubbles penetrating into the liquid nozzle $L_b$. (b) $L_b$ as a function of $M$ and $\mathrm{We}$ for $\mathrm{r}=80$. }
 \label{fig:length_bubble}
\end{figure}

For both the bubbly-jet and flow-blurring regimes, the gas penetrates back into the liquid nozzle. The maximum penetration depth of the gas bubbles, $L_b$, can be measured based on the interfaces, as demonstrated in Fig.~\ref{fig:length_bubble}(a) for the case $\mathrm{r}=80$, $M=1280$, $\mathrm{We}=320$. Figure~\ref{fig:length_bubble}(b) then shows $L_b$ as a function of $M$ for different $\mathrm{We}$. The different regimes can also be easily recognized. For lower $M$ and the airblast-jet regime, $L_b$ is very small. When the two-phase flow transitions to the bubbly-jet or flow-blurring regimes, $L_b$ jumps up, from about 0.5 to 2. There is no distinct change of $L_b$ between the bubbly-jet and flow-blurring regimes. For both regimes, $L_b$ appears to increase approximately linearly with $\log(M)$, as shown in the figure. The transitional dynamic pressure ratio from the airblast-jet regime to the bubbly-jet regime is $M_{tr}=1280$ for $\mathrm{We}=80$, while $M_{tr}=320$ for $\mathrm{We}=320$ and 1280. This indicates that although $M_{tr}$ is generally a function of $\mathrm{We}$, it varies with $\mathrm{We}$ only when $\mathrm{We}$ is not too large. A similar trend is observed for $L_b$. For a given $M$ in the flow-blurring regime, $L_b$ increases as $\mathrm{We}$ increases from 80 to 320, but it varies little when $\mathrm{We}$ further increases from 320 to 1280.

\subsection{Effects of density ratio ($\mathrm{r}$) and gas-to-liquid ratio (GLR)}
\begin{figure}
 \centering{\includegraphics[trim=0.in 0.in 0.in 0.in,clip, width=.99\textwidth]{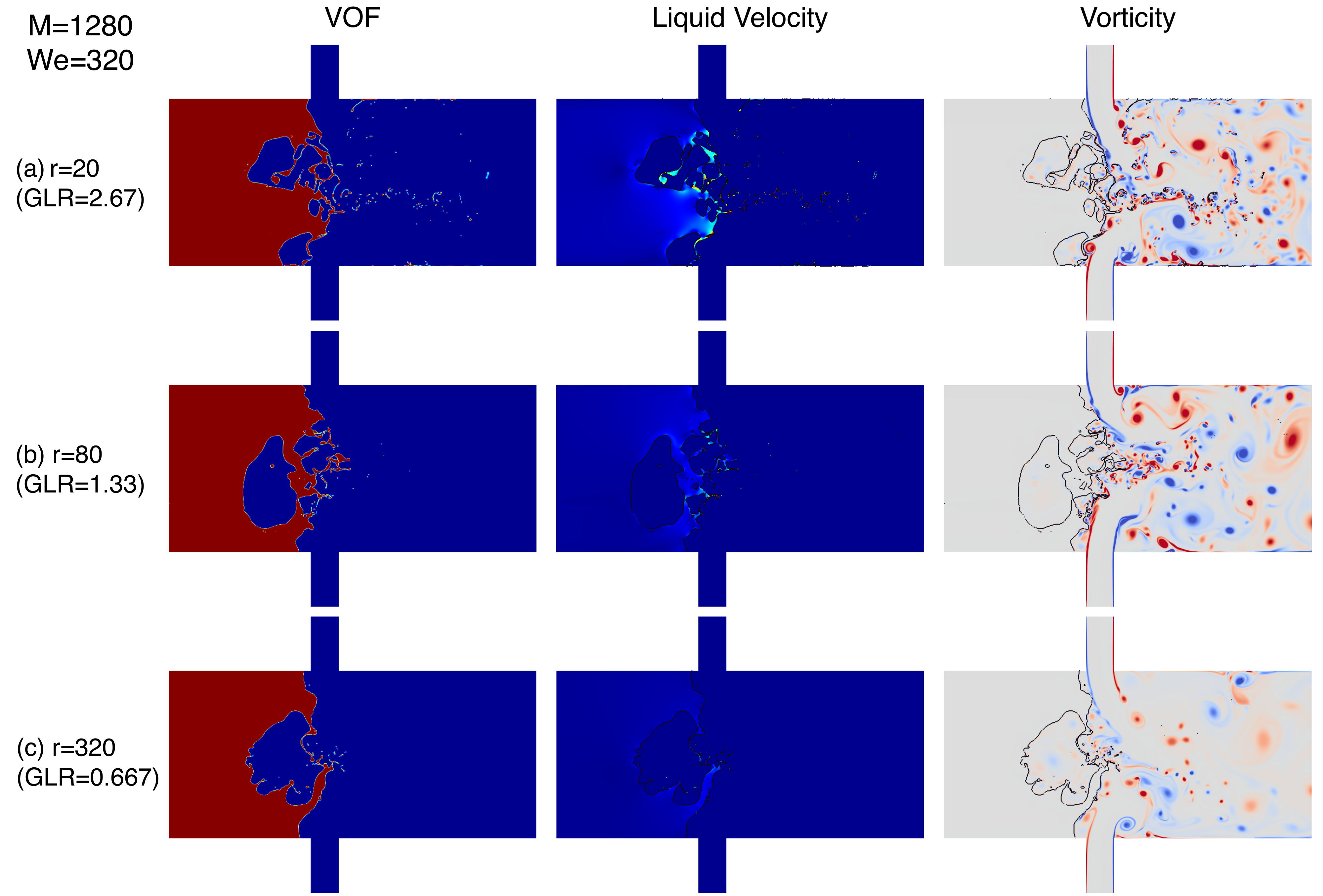}}
 \caption{Representative snapshots of the VOF function ($C$), liquid velocity magnitude ($c|\mathbf{u}|$), and vorticity ($\omega$) for different $\mathrm{r}$ and $\mathrm{GLR}$, while $M=1280$ and $\mathrm{We}=320$ are fixed. The color scales for the liquid velocity and the vorticity are the same as Fig.~\ref{fig:breakup_mode}. }
 \label{fig:effect_r}
\end{figure}

The previous results were all obtained for the same density ratio, \ie $\mathrm{r}=80$, its effect is examined in this section. Figure~\ref{fig:effect_r} shows the results for different $\mathrm{r}$, while $M=1280$ and $\mathrm{We}=320$ are fixed. It can be observed that all three cases belong to the flow-blurring regime, and the gas bubble penetration lengths are also similar, indicating that the effect of $\mathrm{r}$ on the breakup mode transition is less important, compared to $M$ and $\mathrm{We}$. The main difference lies in the number of bubbles formed inside the liquid nozzle. For the case with $\mathrm{r}=320$, only a large bubble is observed. In contrast, for $\mathrm{r}=20$, the large bubble near the center axis is significantly smaller, but there are more small bubbles in the internal two-phase mixing zone.

It is also important to note that as $\mathrm{r}$ increases from 20 to 320, the gas-to-liquid mass flow ratio, $\mathrm{GLR}$, decreases from 2.67 to 0.667. It can be easily shown that $\mathrm{GLR} \sim r^{-1/2}$ when $M$ and $\eta$ are fixed. In previous studies \citep{Qavi_2021f, Murugan_2021f}, $\mathrm{GLR}$ is often used to characterize the flow-blurring regime and the gas bubble penetration depth. However, the results seem to indicate that the effect of $\mathrm{GLR}$ is actually secondary. In previous studies, the flow regime and $L_b$ mainly varied due to the increase in $M$ with $\mathrm{GLR}$, while $\mathrm{r}$ and $\eta$ were fixed. In the present simulation results shown in Fig.~\ref{fig:effect_r}, $\mathrm{GLR}$ is varied by changing $\mathrm{r}$ while keeping $M$ fixed. Then it is clearly seen that $L_b$ varies little with $\mathrm{GLR}$. As long as $M$ is sufficiently high, even a small $\mathrm{GLR}$ can induce a flow-blurring breakup mode.

\section{Effect of contact angle on nozzle wall}
\label{sec:contact_angle}
As can be seen in Fig.~\ref{fig:breakup_mode}, for cases with low $M$, the contact line where the gas, liquid, and solid phases meet is pinned at the two corners between the liquid and top/bottom gas nozzles. In contrast, when $M$ is large, the gas can flow back into the nozzle, and the contact line will move along the liquid nozzle surfaces. Similar contact line motion on the liquid nozzle wall has also been observed in airblast atomization experiments with large $M$ \citep{Heindel_2018a}. In the present simulation, there are no explicit models for moving contact lines \citep{Afkhami_2018a, Qian_2006r}; the movement of the contact line is simply due to the intrinsic numerical slip in the VOF and finite-volume methods. Although a no-slip velocity boundary condition is applied on the solid surface, the velocity at the finite-volume cell center is not zero. If a constant contact angle $\theta$ is invoked on the solid surface boundary, then the reconstructed VOF interface will move along the surface as the value of $C$ in the cell adjacent to the solid surface varies over time.

The mesh size required to accurately resolve the dynamics is significantly smaller than the physical scale of interest in the present problem. Therefore, even for 2D simulations, it is computationally costly to investigate the detailed contact line dynamics near the solid wall, and it is beyond the scope of the present study. However, a parametric study exploring different $\theta$ values has been conducted to examine the leading-order effect of surface wettability on flow-blurring atomization.

\begin{figure}
 \centering{\includegraphics[trim=0.in 0.in 0.in 0.in,clip, width=.99\textwidth]{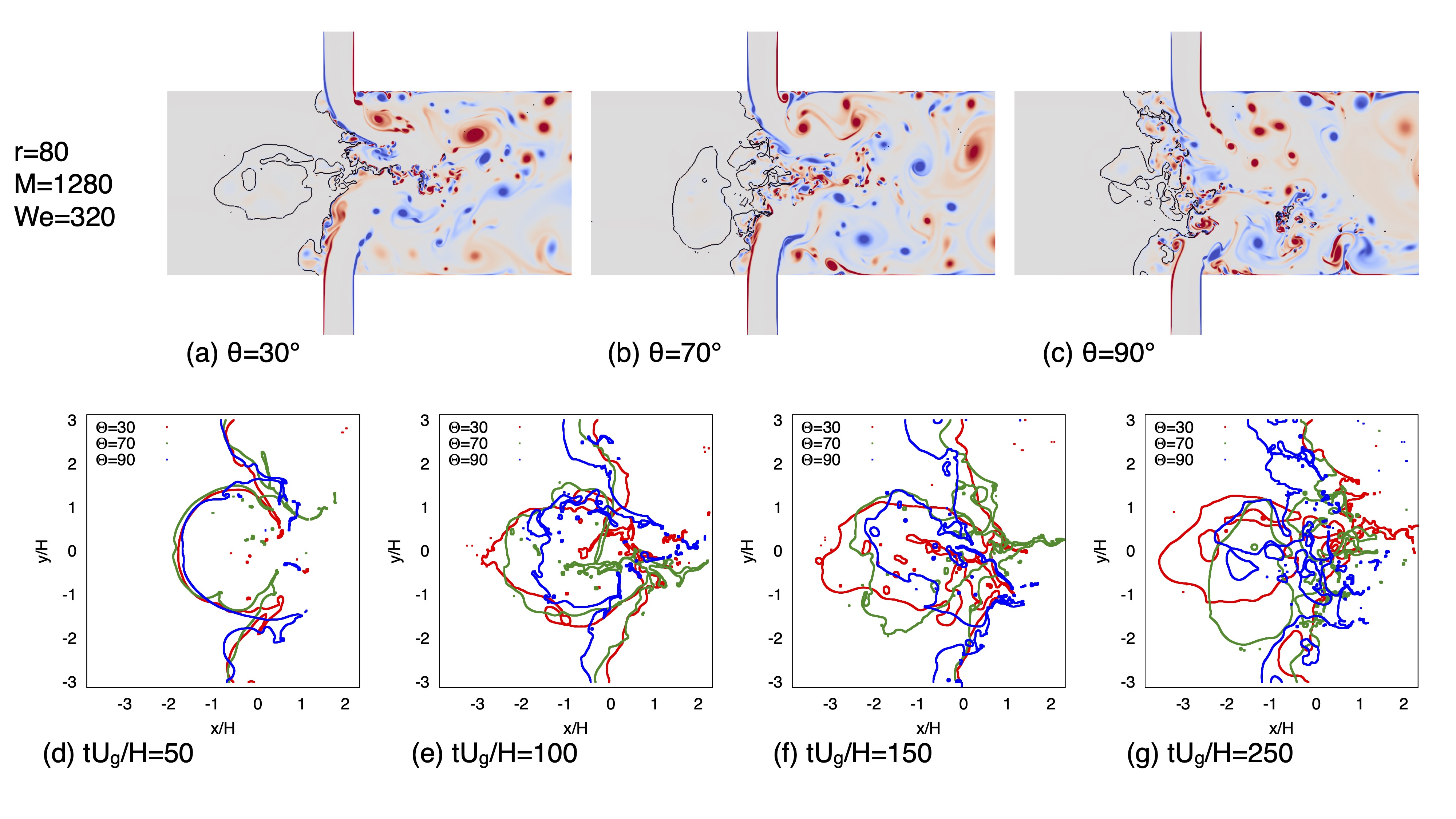}}
 \caption{(a)-(c) Representative snapshots of vorticity ($\omega$) for different contact angles $\theta$ on the liquid nozzle walls, and (d-g) the interfaces at different times. The cases shown here for $r=80$, $M=1280$, and $\mathrm{We}=320$. The color scale for the vorticity is the same as Fig.~\ref{fig:breakup_mode}. }
 \label{fig:contact_angle}
\end{figure}

The time snapshots of vorticity and interfaces for $\theta=30$\textdegree, 70\textdegree, and 90\textdegree\ are shown in Fig.~\ref{fig:contact_angle}. The parameters $r=80$, $M=1280$, and $\mathrm{We}=320$ are identical for all three cases. Despite the wide variation in $\theta$,  all three cases are in the flow-blurring regime due to the high values of $M$ and $\mathrm{We}$, resulting in the formation of the bubbly two-phase mixing zone inside the liquid nozzle. However, the morphology of the bubbly two-phase flows differs significantly, except for the very early stage of development at $tU_g/H=50$ (see Fig.~\ref{fig:contact_angle}(d)). Notably, for the case of $\theta=90$\textdegree, a large bubble does not form near the central axis of the liquid nozzle, unlike the other two cases. The two-phase flow patterns for $\theta=70$\textdegree\ and 30\textdegree\ are more similar, although the size of the primary gas bubble increases as $\theta$ decreases. Additionally, the major bubble is more elongated in the $x$ direction for $\theta=30$\textdegree, compared to $\theta=70$\textdegree.

The different morphologies of the bubbly two-phase zone are a result of modulated interfacial instability, which, in turn, is influenced by the dynamics or mobility of the contact line. When the contact angle $\theta$ is small, the liquid tends to wet the surface of the liquid nozzle, causing the contact line to remain near the corner between the liquid and gas nozzles. It can be observed that the contact line for $\theta=30$ degrees moves within a much smaller spatial range, compared to the other two cases. In the case of $\theta=90$ degrees, the higher mobility of the contact line leads to the formation of gas cavities near the solid surface. As a result, the size of the primary gas bubble near the central axis of the nozzle becomes significantly smaller, and more smaller bubbles are formed. It should be noted that the two-phase flow pattern observed for $\theta=90$ degrees has not been reported in experiments, possibly because most nozzle materials are generally hydrophilic. Nonetheless, the present results suggest that the wettability of the nozzle surfaces does have an influence on the two-phase flow pattern and the liquid breakup dynamics. Therefore, a more comprehensive investigation into the effects of nozzle surface features, such as the contact angle and contact line hysteresis, on flow-blurring atomization is recommended in future studies.

\section{Conclusions}
\label{sec:conc}
In the present study, 2D interface-resolved simulations have been performed to investigate two-phase flows inside a planar flow-blurring atomizer and near the atomizer exit. Parametric simulations have been conducted to examine the effects of the dynamic pressure ratio ($M$), the Weber number ($\mathrm{We}$), and the liquid-to-gas density ratio on the two-phase flow pattern and liquid breakup mode. Four different regimes, including the flapping-jet, airblast-jet, bubbly-jet, and flow-blurring regimes, have been identified. When $\mathrm{We}$ is low, the breakup mode is in the flapping-jet regime. For moderate and high $\mathrm{We}$, the flow can transition from the airblast-jet regime to the bubbly-jet regime and, finally, to the flow-blurring regime as $M$ increases. The present simulations successfully reproduce the flow-blurring atomization mode for sufficiently high $M$ and $\mathrm{We}$, characterized by the formation of a bubbly two-phase  mixing zone inside the liquid nozzle. The high spatial and temporal resolution results reveal the detailed interaction between the bubbly two-phase mixture and the impinging gas jets, leading to the breakup of thin liquid sheets around the bubbles into fine droplets. The gas bubbles penetrate back into the liquid nozzle only in the bubbly-jet and flow-blurring regimes, yet the gas bubble penetration depth ($L_b$) needs to be sufficiently long to achieve a flow-blurring mode. The effect of $\mathrm{We}$ on $L_b$ is less pronounced, in contrast for high $\mathrm{We}$, $L_b$ scales with $\log(M)$. The bubbly-jet regime acts as a transition regime between the more stable airblast-jet and flow-blurring regimes, and thus it is more sensitive to variations in $\mathrm{We}$. With a decrease or increase in $\mathrm{We}$, the two-phase flow for a moderate $M$ may transition to the airblast-jet or flow-blurring regimes. The effects of the density ratio $\mathrm{r}$ and the gas-to-liquid mass ratio $\mathrm{GLR}$ have also been investigated. Although $\mathrm{GLR}$ was widely used in previous studies to characterize the flow-blurring regime, the present results show that its effect is actually secondary compared to $M$ and $\mathrm{We}$. The leading-order effect of liquid nozzle surface wettability is investigated by varying the contact angle on the nozzle walls. While the flow-blurring mode is observed for cases with different contact angles, the bubbly two-phase mixing zone show different morphologies, in particular for the high contact angle $\theta=90$\textdegree.


\acknowledgements

This research was supported by NSF (\#1942324). The High Performance Computing centers at Baylor University and University of South Carolina have provided the computational resources that contributed to the simulation results reported in this paper. The present simulations were performed using the \emph{Gerris} solver, which is made available by Stephane Popinet and other collaborators. The authors also acknowledged Dr.~Udo Fritsching for the helpful discussions.

 \appendix

\section{Grid refinement studies}
\label{sec:grid_refinement}
\begin{figure}
 \centering{\includegraphics[trim=0.in 0.in 0.in 0.in,clip, width=.99\textwidth]{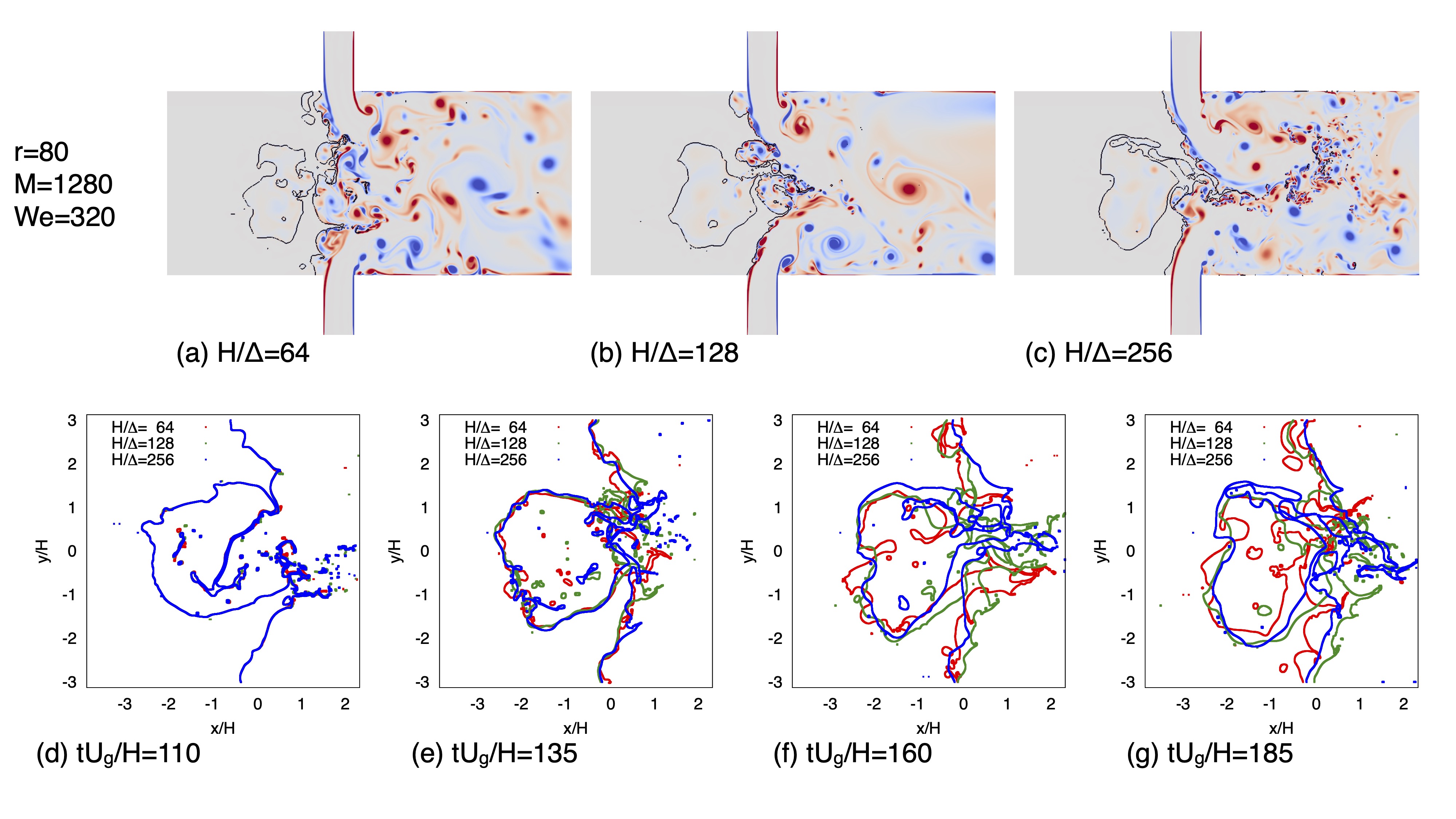}}
 \caption{(a)-(c) Representative snapshots of vorticity ($\omega$) for different mesh resolutions and (d-g) the interfaces at different times. The case shown here for $r=80$, $M=1280$, and $\mathrm{We}=320$. The color scale for the vorticity is the same as Fig.~\ref{fig:breakup_mode}. }
 \label{fig:mesh}
\end{figure}

To examine the effect of mesh resolution on the simulation results, a grid refinement study is conducted using the case with $r=80$, $M=1280$, and $\mathrm{We}=320$. The simulation is initially run with a mesh size of $H/\Delta=128$ until approximately $tU_g/H=110$, as shown in Fig.\ref{fig:mesh}(d). Then, the same snapshot is used to restart the simulations with three different mesh resolutions: $H/\Delta=64$, 128, and 256. The results of the gas-liquid interface and vorticity for the different meshes at a significantly later time, $tU_g/H=182$, are presented in Figs.~\ref{fig:mesh}(a)-(c). The interfaces at different times are shown in Figs.~\ref{fig:mesh}(d)-(g). It can be observed that the interfaces for all three meshes are quite similar. Considering the chaotic nature of the two-phase flows, the agreement between the two finer meshes is very good. The most refined simulations capture small droplets that are missed by the other two meshes. Therefore, although the current mesh size of $H/\Delta=128$ is sufficient to represent the breakup mode and two-phase flow regimes, a finer mesh may be required for future 3D simulations to accurately capture the droplet statistics.

\begin{thebibliography}{46}
\expandafter\ifx\csname natexlab\endcsname\relax\def\natexlab#1{#1}\fi
\expandafter\ifx\csname url\endcsname\relax
  \def\url#1{\texttt{#1}}\fi
\expandafter\ifx\csname urlprefix\endcsname\relax\def\urlprefix{URL }\fi

\bibitem[{Afkhami et~al.(2018)Afkhami, Buongiorno, Guion, Popinet, Saade,
  Scardovelli, and Zaleski}]{Afkhami_2018a}
Afkhami, S., Buongiorno, J., Guion, A., Popinet, S., Saade, Y., Scardovelli,
  R., and Zaleski, S., Transition in a numerical model of contact line dynamics
  and forced dewetting, {\em J.~Comput.~Phys.}, vol.~{\bf 374}, pp. 1061--1093,
  2018.

\bibitem[{Afkhami and Bussmann(2009)}]{Afkhami_2009a}
Afkhami, S. and Bussmann, M., Height functions for applying contact angles to
  3d {VOF} simulations, {\em Int.~J.~Numer.~Meth.~Fluids}, vol.~{\bf 61},
  no.~8, pp. 827--847, 2009.

\bibitem[{Agbaglah et~al.(2017)Agbaglah, Chiodi, and
  Desjardins}]{Agbaglah_2017a}
Agbaglah, G., Chiodi, R., and Desjardins, O., Numerical simulation of the
  initial destabilization of an air-blasted liquid layer, {\em J. Fluid Mech.},
  vol.~{\bf 812}, pp. 1024--1038, 2017.

\bibitem[{Agrawal et~al.(2013)Agrawal, Jiang, Agrawal, and
  Midkiff}]{Agrawal_2013a}
Agrawal, S.R., Jiang, L., Agrawal, A.K., and Midkiff, K.C., High-speed
  visualization of two-phase flow inside a transparent fuel injector, {\em
  {Proc. of the 8th U. S. National Combustion Meeting}}, pp. 070HE--0317, 2013.

\bibitem[{Ates et~al.(2021)Ates, Giovannoni, B{\"u}rkle, Keller, Okraschevski,
  Koch, and Bauer}]{Ates_2021b}
Ates, C., Giovannoni, V., B{\"u}rkle, N., Keller, M., Okraschevski, M., Koch,
  R., and Bauer, H.J., Analysis of the backflow recirculation in flow blurring
  nozzles via lagrangian coherent structures, {\em ICLASS 2021, 15th, Triennial
  International Conference on Liquid Atomization and Spray Systems, Edinburgh,
  UK, 30 Aug.-2 Sept. 2021}, 2021.

\bibitem[{Bozonnet et~al.(2022)Bozonnet, Matas, Balarac, and
  Desjardins}]{Bozonnet_2022a}
Bozonnet, C., Matas, J.P., Balarac, G., and Desjardins, O., Stability of an
  air--water mixing layer: focus on the confinement effect, {\em J.~Fluid
  Mech.}, vol.~{\bf 933}, p. A14, 2022.

\bibitem[{Danh et~al.(2019)Danh, Jiang, and Akinyemi}]{Danh_2019a}
Danh, V., Jiang, L., and Akinyemi, O.S., Investigation of water spray
  characteristics in the near field of a novel swirl burst injector, {\em
  Exp.~Therm Fluid Sci.}, vol.~{\bf 102}, pp. 376--386, 2019.

\bibitem[{Deike et~al.(2016)Deike, Melville, and Popinet}]{Deike_2016a}
Deike, L., Melville, W.K., and Popinet, S., Air entrainment and bubble
  statistics in breaking waves, {\em J. Fluid Mech.}, vol.~{\bf 801}, pp.
  91--129, 2016.

\bibitem[{Delon et~al.(2018)Delon, Cartellier, and Matas}]{Delon_2018a}
Delon, A., Cartellier, A., and Matas, J.P., Flapping instability of a liquid
  jet, {\em Phys.~Rev.~Fluids}, vol.~{\bf 3}, p. 043901, 2018.

\bibitem[{Deshpande et~al.(2015)Deshpande, Gurjar, and
  Trujillo}]{Deshpande_2015o}
Deshpande, S.S., Gurjar, S.R., and Trujillo, M.F., A computational study of an
  atomizing liquid sheet, {\em Phys.~Fluids}, vol.~{\bf 27}, p. 082108, 2015.

\bibitem[{Fisher et~al.(2018)Fisher, Weismiller, Tuttle, and
  Hinnant}]{Fisher_2018a}
Fisher, B.T., Weismiller, M.R., Tuttle, S.G., and Hinnant, K.M., Effects of
  fluid properties on spray characteristics of a flow-blurring atomizer, {\em
  J.~Eng.~Gas.~Turbines Power-Trans.~{ASME}}, vol.~{\bf 140}, p. 041511, 2018.

\bibitem[{Francois et~al.(2006)Francois, Cummins, Dendy, Kothe, Sicilian, and
  Williams}]{Francois_2006a}
Francois, M.M., Cummins, S.J., Dendy, E.D., Kothe, D.B., Sicilian, J.M., and
  Williams, M.W., A balanced-force algorithm for continuous and sharp
  interfacial surface tension models within a volume tracking framework, {\em
  J.~Comput.~Phys.}, vol.~{\bf 213}, pp. 141--173, 2006.

\bibitem[{Fuster et~al.(2013)Fuster, Matas, Marty, Popinet, J., Cartellier, and
  Zaleski}]{Fuster_2013a}
Fuster, D., Matas, J.P., Marty, S., Popinet, S., J., H., Cartellier, A., and
  Zaleski, S., Instability regimes in the primary breakup region of planar
  coflowing sheets, {\em J. Fluid Mech}, vol.~{\bf 736}, pp. 150--176, 2013.

\bibitem[{Ganan-Calvo(2005)}]{Ganan-Calvo_2005a}
Ganan-Calvo, A.M., Enhanced liquid atomization: From flow-focusing to
  flow-blurring, {\em Appl.~Phys.~Lett.}, vol.~{\bf 86}, p. 214101, 2005.

\bibitem[{Heindel(2018)}]{Heindel_2018a}
Heindel, T., X-ray imaging techniques to quantify spray characteristics in the
  near field, {\em Atomization Spray}, vol.~{\bf 28}, pp. 1029--1059, 2018.

\bibitem[{Herrmann(2010)}]{Herrmann_2010a}
Herrmann, M., A parallel eulerian interface tracking/lagrangian point particle
  multi-scale coupling procedure, {\em J.~Comput.~Phys.}, vol.~{\bf 229}, pp.
  745--759, 2010.

\bibitem[{Jaber et~al.(2020)Jaber, Kourmatzis, and Masri}]{Jaber_2020a}
Jaber, O.J., Kourmatzis, A., and Masri, A.R., Characterization of flow-focusing
  and flow-blurring modes of atomization, {\em Energy Fuels}, vol.~{\bf 35},
  pp. 7144--7155, 2020.

\bibitem[{Jackiw and Ashgriz(2022)}]{Jackiw_2022a}
Jackiw, I.M. and Ashgriz, N., Prediction of the droplet size distribution in
  aerodynamic droplet breakup, {\em J.~Fluid Mech.}, vol.~{\bf 940}, p. A17,
  2022.

\bibitem[{Jiang and Ling(2021)}]{Jiang_2021a}
Jiang, D. and Ling, Y., Impact of inlet gas turbulence on the formation,
  development and breakup of interfacial waves in a two-phase mixing layer,
  {\em J.~Fluid Mech.}, vol.~{\bf 921}, p. A15, 2021.

\bibitem[{Jiang and Agrawal(2015{\natexlab{a}})}]{Jiang_2015b}
Jiang, L. and Agrawal, A.K., Investigation of glycerol atomization in the
  near-field of a flow-blurring injector using time-resolved {PIV} and
  high-speed visualization, {\em Flow Turbul.~Combust.}, vol.~{\bf 94}, pp.
  323--338, 2015{\natexlab{a}}.

\bibitem[{Jiang and Agrawal(2015{\natexlab{b}})}]{Jiang_2015a}
Jiang, L. and Agrawal, A.K., Spray features in the near field of a
  flow-blurring injector investigated by high-speed visualization and
  time-resolved {PIV}, {\em Exp.~Fluids}, vol.~{\bf 56}, p. 103,
  2015{\natexlab{b}}.

\bibitem[{Khan et~al.(2019)Khan, Gadgil, and Kumar}]{Khan_2019a}
Khan, M.A., Gadgil, H., and Kumar, S., Influence of liquid properties on
  atomization characteristics of flow-blurring injector at ultra-low flow
  rates, {\em Energy}, vol.~{\bf 171}, pp. 1--13, 2019.

\bibitem[{Lasheras and Hopfinger(2000)}]{Lasheras_2000a}
Lasheras, J.C. and Hopfinger, E.J., Liquid jet instability and atomization in a
  coaxial gas stream, {\em Annu. Rev. Fluid Mech.}, vol.~{\bf 32}, pp.
  275--308, 2000.

\bibitem[{Lasheras et~al.(1998)Lasheras, Villermaux, and
  Hopfinger}]{Lasheras_1998a}
Lasheras, J.C., Villermaux, E., and Hopfinger, E.J., Break-up and atomization
  of a round water jet by a high-speed annular air jet, {\em J.~Fluid Mech.},
  vol.~{\bf 357}, pp. 351--379, 1998.

\bibitem[{Lefebvre(1980)}]{Lefebvre_1980a}
Lefebvre, A.H., Airblast atomization, {\em Prog.~Energ.~Combust.~Sci.},
  vol.~{\bf 6}, pp. 233--261, 1980.

\bibitem[{Ling et~al.(2019)Ling, Fuster, Tryggvasson, and Zaleski}]{Ling_2019a}
Ling, Y., Fuster, D., Tryggvasson, G., and Zaleski, S., A two-phase mixing
  layer between parallel gas and liquid streams: multiphase turbulence
  statistics and influence of interfacial instability, {\em J.~Fluid Mech.},
  vol.~{\bf 859}, pp. 268--307, 2019.

\bibitem[{Ling et~al.(2017)Ling, Fuster, Zaleski, and Tryggvason}]{Ling_2017a}
Ling, Y., Fuster, D., Zaleski, S., and Tryggvason, G., Spray formation in a
  quasiplanar gas-liquid mixing layer at moderate density ratios: A numerical
  closeup, {\em Phys.~Rev.~Fluids}, vol.~{\bf 2}, p. 014005, 2017.

\bibitem[{Ling and Mahmood(2023)}]{Ling_2023a}
Ling, Y. and Mahmood, T., 2023. Detailed numerical investigation of the drop
  aerobreakup in the bag breakup regime, arXiv:2305.00124.

\bibitem[{Marmottant and Villermaux(2004)}]{Marmottant_2004a}
Marmottant, P. and Villermaux, E., On spray formation, {\em J.~Fluid Mech.},
  vol.~{\bf 498}, pp. 73--111, 2004.

\bibitem[{Matas et~al.(2011)Matas, Marty, and Cartellier}]{Matas_2011a}
Matas, J.P., Marty, S., and Cartellier, A., Experimental and analytical study
  of the shear instability of a gas-liquid mixing layer, {\em Phys.~Fluids},
  vol.~{\bf 23}, p. 094112, 2011.

\bibitem[{Murugan and Kolhe(2021)}]{Murugan_2021f}
Murugan, R. and Kolhe, P.S., Experimental investigation into flow blurring
  atomization, {\em Exp.~Therm.~Fluid Sci.}, vol.~{\bf 120}, p. 110240, 2021.

\bibitem[{Murugan et~al.(2020)Murugan, Kolhe, and Sahu}]{Murugan_2020j}
Murugan, R., Kolhe, P.S., and Sahu, K.C., A combined experimental and
  computational study of flow-blurring atomization in a twin-fluid atomizer,
  {\em Eur.~J.~Mech.~B/Fluids}, vol.~{\bf 84}, pp. 528--541, 2020.

\bibitem[{Neel and Villermaux(2018)}]{Neel_2018r}
Neel, B. and Villermaux, E., The spontaneous puncture of thick liquid films,
  {\em J.~Fluid Mech.}, vol.~{\bf 838}, pp. 192--221, 2018.

\bibitem[{Opfer et~al.(2014)Opfer, Roisman, Venzmer, Klostermann, and
  Tropea}]{Opfer_2014a}
Opfer, L., Roisman, I.V., Venzmer, J., Klostermann, M., and Tropea, C.,
  Droplet-air collision dynamics: Evolution of the film thickness, {\em
  Phys.~Rev.~E}, vol.~{\bf 89}, p. 013023, 2014.

\bibitem[{Oron et~al.(1997)Oron, Davis, and Bankoff}]{Oron_1997o}
Oron, A., Davis, S.H., and Bankoff, S.G., Long-scale evolution of thin liquid
  films, {\em Rev.~Mod.~Phys}, vol.~{\bf 69}, p. 931, 1997.

\bibitem[{Popinet(2003)}]{Popinet_2003a}
Popinet, S., Gerris: {A} tree-based adaptive solver for the incompressible
  {E}uler equations in complex geometries, {\em J.~Comput.~Phys.}, vol.~{\bf
  190}, pp. 572--600, 2003.

\bibitem[{Popinet(2009)}]{Popinet_2009a}
Popinet, S., An accurate adaptive solver for surface-tension-driven interfacial
  flows, {\em J.~Comput.~Phys.}, vol.~{\bf 228}, no.~16, pp. 5838--5866, 2009.

\bibitem[{Qavi et~al.(2021)Qavi, Jiang, and Akinyemi}]{Qavi_2021f}
Qavi, I., Jiang, L., and Akinyemi, O.S., Near-field spray characterization of a
  high-viscosity alternative jet fuel blend c-3 using a flow blurring injector,
  {\em Fuel}, vol.~{\bf 293}, p. 120350, 2021.

\bibitem[{Qian et~al.(2006)Qian, Wang, and Sheng}]{Qian_2006r}
Qian, T., Wang, X.P., and Sheng, P., A variational approach to moving contact
  line hydrodynamics, {\em J.~Fluid Mech.}, vol.~{\bf 564}, pp. 333--360, 2006.

\bibitem[{Sallevelt et~al.(2015)Sallevelt, Pozarlik, and
  Brem}]{Sallevelt_2015i}
Sallevelt, J.L.H.P., Pozarlik, A.K., and Brem, G., Characterization of viscous
  biofuel sprays using digital imaging in the near field region, {\em
  Appl.~Energy}, vol.~{\bf 147}, pp. 161--175, 2015.

\bibitem[{Scardovelli and Zaleski(1999)}]{Scardovelli_1999a}
Scardovelli, R. and Zaleski, S., Direct numerical simulation of free-surface
  and interfacial flow, {\em Annu.~Rev.~Fluid Mech.}, vol.~{\bf 31}, pp.
  567--603, 1999.

\bibitem[{Shinjo and Umemura(2010)}]{Shinjo_2010a}
Shinjo, J. and Umemura, A., Simulation of liquid jet primary breakup: Dynamics
  of ligament and droplet formation, {\em Int.~J.~Multiphase Flow}, vol.~{\bf
  36}, pp. 513--532, 2010.

\bibitem[{Simmons and Agrawal(2010)}]{Simmons_2010h}
Simmons, B.M. and Agrawal, A., Spray characteristics of a flow-blurring
  atomizer, {\em Atomization Spray}, vol.~{\bf 20}, pp. 821--835, 2010.

\bibitem[{Tang et~al.(2023)Tang, Adcock, and Mostert}]{Tang_2023a}
Tang, K., Adcock, T., and Mostert, W., Bag film breakup of droplets in uniform
  airflows, {\em J.~Fluid Mech.}, vol.~{\bf 970}, p.~A9, 2023.

\bibitem[{Zhang et~al.(2019)Zhang, Ling, Tsai, Wang, Popinet, and
  Zaleski}]{Zhang_2019b}
Zhang, B., Ling, Y., Tsai, P.H., Wang, A.B., Popinet, S., and Zaleski, S.,
  Short-term oscillation and falling dynamics for a water drop dripping in
  quiescent air, {\em Phys.~Rev.~Fluids}, vol.~{\bf 4}, p. 123604, 2019.

\bibitem[{Zhao et~al.(2019)Zhao, Ning, Lu, and Wang}]{Zhao_2019a}
Zhao, J., Ning, Z., Lu, M., and Wang, G., Numerical simulation of flow focusing
  pattern and morphological changes in two-phase flow inside nozzle, {\em
  Chin.~J.~Chem.~Eng.}, vol.~{\bf 27}, pp. 63--71, 2019.

\end{thebibliography}

\end{document}